\documentclass[aps, preprint, superscriptaddress, nofootinbib]{revtex4}

\usepackage{amsfonts, amsbsy, amssymb, amsmath, color,  float}
\usepackage[pdftex]{graphicx}

\newcommand{\rmd}{{\rm d}}
\newcommand{\deriv}[2]{{\frac{\rmd {#1}}{\rmd {#2}}}}

\newcommand{\pd}[2]{\frac{\partial {#1}}{\partial {#2}}}

\newcommand{\calK}{\mathcal{K}}
\newcommand{\calH}{\mathcal{H}}
\newcommand{\calE}{\mathcal{E}}

\newcommand{\calM}{\mathcal{M}}
\newcommand{\calP}{\mathcal{P}}
\newcommand{\calV}{\mathcal{V}}
\newcommand{\calI}{\mathcal{I}}
\newcommand{\calJ}{\mathcal{J}}
\newcommand{\bbP}{\mathbb{P}}

\newcommand{\bq}{\boldsymbol{q}}
\newcommand{\bz}{\boldsymbol{z}}
\newcommand{\betab}{\bar{\beta}}
\newcommand{\omegab}{\bar{\omega}}
\newcommand{\qb}{\bar{q}}

\newcommand{\rb}{\bar{r}}
\newcommand{\bqb}{\bar{\boldsymbol{q}}}
\newcommand{\bpb}{\bar{\boldsymbol{p}}}
\newcommand{\bpib}{\bar{\boldsymbol{\pi}}}

\newcommand{\kB}{k_{\text{B}}}

\newcommand{\sbar}{\bar{s}}

\newcommand{\Int}[3]{\int_{#2}^{#3}\rmd {#1} \;}

\begin{document}


\title{Phase space structure and dynamics for the Hamiltonian isokinetic thermostat}



\author{Peter Collins}
\affiliation{School of Mathematics\\University of Bristol\\Bristol BS8 1TW\\United Kingdom}

\author{Gregory S. Ezra}
\email[]{gse1@cornell.edu}
\affiliation{Department of Chemistry and Chemical Biology\\
Baker Laboratory\\
Cornell University\\
Ithaca, NY 14853\\USA}

\author{Stephen Wiggins}
\email[]{stephen.wiggins@mac.com}
\affiliation{School of Mathematics\\University of Bristol\\Bristol BS8 1TW\\United Kingdom}

\date{\today}

\begin{abstract}

We investigate the phase space structure and
dynamics of a Hamiltonian isokinetic thermostat, for which 
ergodic thermostat trajectories at fixed (zero) energy generate
a canonical distribution in configuration space.
Model potentials studied consist of
a single bistable mode plus transverse harmonic modes.  
Interpreting the bistable mode as a reaction (isomerization) coordinate, we 
establish connections with the theory of unimolecular reaction rates, 
in particular the formulation of isomerization rates in terms
of gap times. The distribution
of gap times (or associated lifetimes) for a microcanonical ensemble initiated on the dividing surface is
of great dynamical significance; an exponential lifetime
distribution is usually taken to be an indicator of `statistical' behavior.
Moreover, comparison of the magnitude of the phase space volume swept out by reactive
trajectories as they pass through the reactant region with the total 
phase space volume (classical density of states) for the reactant region
provides a necessary condition for ergodic dynamics.
We compute gap times, associated lifetime distributions, mean gap times, reactive
fluxes, reactive  volumes and total reactant phase space volumes for model systems with 3 
degrees of freedom,
at three different temperatures.
At all three temperatures, the necessary condition for ergodicity is approximately satisfied.
At high temperatures 
a non-exponential lifetime distribution is found, while at low temperatures 
the lifetime is more nearly exponential.  The degree of exponentiality of the
lifetime distribution is quantified by computing the information entropy deficit 
with respect to pure exponential decay.
The efficacy of the Hamiltonian isokinetic thermostat is examined by computing
coordinate distributions averaged over single long trajectories initiated
on the dividing surface.

\end{abstract}


\maketitle



\newpage

\section{Introduction}
\label{sec:introduction}

Deterministic thermostats are widely used to simulate equilibrium physical systems
described by ensembles other than microcanonical (constant energy and volume, $(E, V)$),  such as constant
temperature-volume $(T, V)$ or temperature-pressure $(T, p)$
\cite{Nose91,Morriss98,Hoover04,Leimkuhler04a,Hunenberger05,Bond07},
or for simulations of nonequilibrium phenomena
\cite{Evans90,Hoover91,Mundy00,RomeroBastida06,Hoover07a,Jepps10}.

Deterministic thermostats
are obtained by augmenting the phase space variables of the physical system of interest
with a set of additional variables
whose role is to alter the standard Hamiltonian system dynamics in such a way that
a suitable invariant measure in the system phase space is preserved
\cite{Nose84,Nose91,Hoover85}.
For example, in the familiar Nos\'{e}-Hoover (NH) thermostat \cite{Nose84,Hoover85} the
exact dynamics preserves both
an extended energy and a suitable invariant measure, ensuring that, \emph{provided}
the extended system dynamics
is effectively ergodic on the timescale of the simulation, the physical system will sample its phase space
according to the canonical (constant $T$) measure.

Extended system thermostat dynamics can be either Hamiltonian \cite{Dettmann96,Dettmann97,Dettmann99,Bond99}
or non-Hamiltonian \cite{Evans90,Tuckerman99,Tuckerman01,Sergi01,Sergi03,Ezra04,Tarasov05,Sergi07,Sergi10}.
For example,  NH dynamics is non-Hamiltonian \cite{Hoover85}.
(The NH system is however
conformally symplectic \cite{Wojtkowski98,Choquard98}, meaning that the dynamics can be rendered Hamitonian by a
coordinate-dependent scaling of the vector field \cite{Wojtkowski98}.
This is not true for other non-Hamiltonian
thermostats such as Nos\'{e}-Hoover chains \cite{Martyna92} or 
Bulgac-Kusnezov global demons \cite{Kuznezov90}.)

There has been considerable interest in the formulation of
Hamiltonian deterministic thermostats such as the Nos\'{e}-Poincar\'{e} system \cite{Bond99}.
In this  approach,
the extended Hamiltonian for the physical system plus thermostat variables
incorporates a coordinate-dependent time scaling of Poincar\'{e}-Sundman type \cite{Bond98,Benest02}.
Restricting the dynamics to a fixed value (zero) of the extended Hamiltonian results in the system variables
sampling their phase space according to (for example) the canonical density \cite{Bond99} (again, subject to the
assumption of ergodicity).
An important motivation for the introduction of Hamiltonian thermostats is the possibility of using
symplectic integration algorithms to integrate trajectories \cite{SanzSerna94,Hairer02,Leimkuhler04a}.

As already indicated, a fundamental question concerning deterministic thermostats has to do with the
effective ergodicity of the dynamics on the timescale of the simulation.
If the dynamics is not effectively ergodic, then
trajectory simulations will not generate the correct invariant measure \cite{Cornfeld82,Sturman06}.
It has long been recognized, for example,  that the dynamical system
consisting of a single harmonic oscillator degree of freedom coupled to the NH thermostat
variable is not ergodic \cite{Hoover85} (see also refs
\onlinecite{Golo04,Watanabe07,Watanabe07a}); in fact, for typical parameter values
the system phase space exhibits extensive persistence of invariant tori (quasiperiodic motion) 
\cite{Legoll07,Legoll09}.
A large amount of effort has  been expended in attempts
to design thermostats exhibiting dynamics more ergodic than the basic NH system
\cite{Martyna92,Leimkuhler04a,Hunenberger05,Leimkuhler05a}.
(For a careful discussion of the question of ergodicity for `typical' interatomic potentials, see
\cite{Tupper05}; for related fundamental questions in molecular dynamics, 
see also ref.\ \cite{Skeel09}.)

The question of ergodicity in thermostats is conceptually closely related  to the problem of
statistical versus nonstatistical behavior in the (classical) theory of unimolecular reaction
rates \cite{Robinson72,Gilbert90,Baer96,Forst03}.
Broadly speaking, in this case one would like to know whether a molecule will
behave according to a statistical model such as RRKM theory, or 
whether it will exhibit significant deviations
from such a theory, ascribable 
to nonstatistical dynamics \cite{Brumer88,Rice81,DeLeon81,Rice96,Carpenter05,Ezra09a}.
Such `nonstatisticality', which
can arise from a number of dynamical effects,
can be considered to be 
analogous to the failure of ergodicity in deterministic thermostats.

In recent years there have been significant theoretical and computational advances in the application
of dynamical systems theory \cite{Wiggins90,Wiggins92,Wiggins94}
to study reaction dynamics and phase space structure in multimode models of molecular systems,
and to probe the
dynamical origins of nonstatistical behavior
\cite{wwju,Komatsuzaki02,Komatsuzaki05,Jaffe05,Wiesenfeld05,WaalkensSchubertWiggins08}.
The fundamental chemical concept
of the \emph{transition state}, defined  as a surface
of no return in phase space, has been successfully and rigorously generalized from the well-established
2 degree of freedom case  \cite{Pechukas81}
to systems with $N \geq 3$ degrees of freedom
\cite{wwju,Komatsuzaki02,Komatsuzaki05,Jaffe05,Wiesenfeld05,WaalkensSchubertWiggins08}.
Moreover, dynamical indicators exist (determination of reactive phase space volume,
behavior of the reactive flux) to diagnose nonstatistical behavior 
(see, for example, Ref.\ \cite{DeLeon81,Ezra09a,Lourderaj09}).

Nevertheless, relatively little work has been done applying the powerful techniques 
from dynamical system theory
\cite{Wiggins90,Wiggins92,Wiggins94}
to study the phase space structure and dynamics of deterministic thermostats.
Hoover and coworkers have investigated the
fractal nature of various phase space structures for equilibrium \cite{Posch86}
and nonequilibrium \cite{Posch97,Hoover98b,Hoover01a} versions of the NH thermostat.
Using a Hamiltonian formulation of
the isokinetic thermostat \cite{Dettmann96} (cf.\ Section \ref{sec:isokinetic_thermostat} of this paper),
Morriss and Dettmann \cite{Morriss98} have mapped the dynamics of the isokinetically
thermostatted Lorentz gas onto the
geodesic motion of a free particle on a particular Riemannian manifold.
Leimkuhler and Sweet have used Poincar\'{e} surfaces of section and dynamical frequency analysis
in their analysis of optimal coupling constants for NH and related thermostats \cite{Leimkuhler05a},
while Legoll
and coworkers have applied KAM theory to rigorously prove the existence of invariant tori in the
NH thermostatted harmonic oscillator \cite{Legoll07,Legoll09}.
D'Alessandro, Tenenbaum and Amadei have computed Lyapunov exponents for 
a thermostatted united-atom model of the butane molecule \cite{DAlessandro02}; both Nos\'{e}-Hoover
and isokinetic Gaussian thermostats were considered.

Thermostatted systems present several difficulties for
detailed studies from a phase space perspective.  First, such systems usually have
high dimensionality; for example, a NH chain thermostatted system has at least 3 degrees of freedom.
For such a system the Poincar\'{e} surface of section analysis, standard for 2 degrees of freedom, does not afford the advantage of dimensional reduction and global visualization since the surface of section is four dimensional for three degrees of freedom. Second, there is a lack of readily computable diagnostics that can be used to  establish
ergodicity \cite{Sturman06}, other than comparison of coordinate distributions obtained using the given
thermostat with those obtained using other methods.

In the present paper we analyze the Hamiltonian isokinetic thermostat \cite{Morriss98}
from the perspective of reaction rate theory.
Although not as widely used as the NH thermostat and its many variants,
the non-Hamiltonian version of the isokinetic thermostat has been developed and applied to
several problems of chemical interest by Minary et al.\ \cite{Minary03a,Minary03b}.
In this thermostat, the particle momenta are subject to a nonholonomic constraint
that keeps the kinetic energy constant.  The resulting dynamics
generates a canonical (constant tenperature) distribution in configuration space \cite{Morriss98}.
In this work we investigate a slightly modified version
of the Hamiltonian isokinetic thermostat given by Litniewski \cite{Litniewski93} 
and Morishita \cite{Morishita03}.

The structure of the present paper is as follows:
Section \ref{sec:isokinetic_thermostat} reviews the 
Hamiltonian formulation of the isokinetic thermostat.  The non-Hamiltonian
equations of motion for a Hamiltonian system subject to the isokinetic constraint
correspond to Hamiltonian dynamics at zero energy under an effective Hamiltonian whose
potential is obtained from the physical potential by exponentiation.
The model Hamiltonians analyzed in the present paper are introduced
in Section \ref{sec:hamiltonians}.  In these systems, the physical potential
describes $n$ uncoupled oscillators, $n-1$ harmonic modes together with
a bistable thermalizing \cite{Minary03a,Minary03b} or isomerization
coordinate.  For these Hamiltonians the physical potential exhibits a saddle of index one,
as for the case of a bistable reaction profile coupled to one or more
transverse confining modes.
In Section \ref{sec:hamiltonian_nonhamiltonian} we briefly review
earlier results \cite{Ezra09} establishing
that the isokinetic Hamiltonian dynamical system corresponding to
the physical Hamiltonian exhibits a
normally hyperbolic invariant manifold
(NHIM) associated with the saddle \cite{Wiggins94}, 
at least for a limited range of
energies above that of the saddle.
The phase space formulation of unimolecular reaction rate theory in terms of
the gap time \cite{Slater56,Thiele62,Dumont86,Ezra09a} 
and related concepts is discussed in Sect.\ \ref{sec:unimolecular}.
The classical spectral theorem 
\cite{Brumer80,Pollak81,Binney85,Meyer86,WaalkensBurbanksWiggins05,WaalkensBurbanksWiggins05c}
provides a relation 
between the distribution of gap times and the phase space volume occupied
by reactive phase points; unless the measure of the region swept out
by isomerizing trajectories equals that of the energy shell, the system
cannot  be ergodic.  This necessary condition for ergodicity
establishes a connection between the concepts of reaction rate theory and
the properties of the Hamiltonian isokinetic thermostat.
Numerical results on 3 and 4 degree of freedom Hamiltonian isokinetic thermostats are reported
in Section \ref{sec:numerical_results}.
Section \ref{sec:summary} concludes.  Analytical results for the density of states
for exponentiated harmonic potential are given in Appendix \ref{sec:dos}, while
details of the phase space sampling procedures used in our numerical work
appear in Appendix \ref{sec:sampling}.

A brief discussion of the theoretical framework developed here
was given in ref.\ \onlinecite{Ezra09}.

\newpage

\section{The Hamiltonian Isokinetic Thermostat}
\label{sec:isokinetic_thermostat}

The role of the isokinetic thermostat is to dynamically
constrain the kinetic energy of a simulated system to have a constant value
proportional to $k_{\text{B}} T$, where $T$ is the desired temperature and $k_{\text{B}}$ is Boltzmann's
constant.  Non-Hamiltonian equations of motion for the isokinetic thermostat
have been obtained by applying Gauss' principle of least constraint \cite{Evans90,Hoover91,Morriss98},
which is the appropriate dynamical principle for nonholonomically constrained systems \cite{Evans83,Evans90}.

Provided the underlying dynamics is effectively ergodic on the timescale of the
simulation, a trajectory of the non-Hamiltonian isokinetic thermostat
samples configuration space according to the invariant measure associated with the
equilibrium canonical ensemble \cite{Evans83a,Morriss98}.  Algorithms for
the non-Hamiltonan isokinetic thermostat
have been developed and applied to a variety of systems by Minary,
Martyna and Tuckerman \cite{Minary03a,Minary03b}.

A Hamiltonian formulation of the isokinetic thermostat has been given by
Dettmann and Morriss \cite{Dettmann96,Morriss98}.  The Hamiltonian used in the Dettmann-Morriss approach
incorporates a coordinate-dependent scaling of time \cite{Szebehely67,Benest02},
and the physically relevant dynamics is restricted to the
zero energy hypersurface.  A  noncanonical tranformation of variables 
then leads to a set of dynamical equations
equivalent to the non-Hamiltonian version.
An important motivation for the development of  Hamiltonian versions  of the isokinetic
thermostat is the possibility of using symplectic integration algorithms to ensure
qualitatively correct behavior of integrated trajectories over long times \cite{SanzSerna94,Hairer02}.

In this section we briefly review the Hamiltonian formulation of the isokinetic thermostat
(for more detailed discussion, see \cite{Ezra09}.)
As our focus here is on systems with $n \geq 3$ degrees of freedom (DoF),
we  discuss a Hamiltonian version of the isokinetic thermostat that  generates
the invariant measure associated with the configurational canonical ensemble, while
at the same time allowing use of the simplest Verlet-type \cite{Verlet67} symplectic integration algorithm
\cite{Litniewski93,Morishita03}.

\subsection{Hamiltonian isokinetic thermostat}

The physical Hamiltonian of interest is assumed to have the standard form
\begin{equation}
\label{ham_1}
H(q, p) = \frac{1}{2} p^2 + \Phi(q),
\end{equation}
where $(q, p) \in \mathbb{R}^n \times \mathbb{R}^n$, $\Phi(q)$ is the potential energy, and we
set all masses equal to unity for simplicity.
Corresponding physical Hamiltonian equations of motion are
\begin{subequations}
\label{ham_eq}
\begin{align}
\dot{q} & = +\pd{H}{p} \\
\dot{p} &= - \pd{H}{q}.
\end{align}
\end{subequations}

 For $n \geqslant 3$ degrees of freedom, we
    define a Hamiltonian $\calH = \calH (\pi, q)$ as \cite{Litniewski93,Morishita03}
    \begin{equation}
      \label{sp_7}
      \calH =  \frac{1}{2}\pi^2 - \frac{\nu}{ 2 \betab} \, e^{-2 \betab \Phi},
  \end{equation}
  where $\pi \in \mathbb{R}^n$ and  $\betab$ and $\nu$ are parameters to be determined.
  The associated Hamiltonian equations of motion are
    \begin{subequations}
    \label{sp_6p}
    \begin{align}
    {q}^\prime & = + \, \pd{\calH}{\pi} = \pi\\
    {\pi}^\prime & = -\, \pd{\calH}{q} = -\nu \Phi_q \, e^{-2 \betab \Phi},
    \end{align}
  \end{subequations}
  where the time derivative is denoted by a prime (${}^\prime$) to indicate that
  the derivative is actually taken with respect to a suitably scaled time variable 
  (see \cite{Morriss98,Ezra09}).
    As the kinetic energy term in \eqref{sp_7} is coordinate-independent,
    a basic Verlet-type symplectic integrator can be used to integrate
    Hamilton's equations for $\calH$ \cite{Verlet67,SanzSerna94,Hairer02,Leimkuhler04a}.
  (Higher-order symplectic algorithms can of course also be used \cite{Hairer02,Leimkuhler04a}.)

  Making a \emph{noncanonical} change of coordinates from variable $(\pi, q)$ to physical variables $(p, q)$,
  $(\pi, q) \mapsto (p, q)$, where
  \begin{equation}
  \pi = e^{-\betab \Phi} p
  \end{equation}
  the Hamiltonian function $\calH$ becomes
  \begin{equation}
  \calH = e^{-2 \betab \Phi} \frac{1}{2} \left[ p^2 - \frac{\nu}{\betab} \right].
  \end{equation}
  Setting $\calH = 0$ then automatically enforces an \emph{isokinetic constraint} in terms of
  the momentum variables $p$ \cite{Morriss98}
  \begin{equation}
  \frac{p^2}{2} =  \frac{\nu}{ 2 \betab}.
  \end{equation}

  For trajectories run at $\calH=0$, and only for this value, we have the
  invariant volume element  \cite{Morriss98}
  \begin{subequations}
  \begin{align}
  \rmd V & = \rmd^n \pi \rmd^n q \, \delta(\calH)  \\
    & = \rmd^n p \rmd^n q \, e^{-n\betab\Phi} \delta(\calH) \\
    & =  \rmd^n p \rmd^n q \, e^{-n\betab\Phi}
    \delta\left[\tfrac{1}{2}e^{-2\betab\Phi}\left(p^2 - \frac{\nu}{\betab} \right)\right]\\
    & =  \rmd^n p \rmd^n q \, 2e^{-(n-2)\betab\Phi} \delta\left[p^2 - \frac{\nu}{\betab} \right].
    \end{align}
    \end{subequations}

  We therefore set ($\beta \equiv 1/k_{\text{B}} T$ as usual)
  \begin{equation}
  \label{betab_1}
  \betab = \frac{1}{(n-2)k_{\text{B}} T} = \frac{\beta}{(n-2)}\,,
  \end{equation}
  to ensure that the invariant measure is proportional to the canonical density (recall $n>2$) 
  \begin{equation}
  \rho(q) \propto \exp[-\Phi(q)/k_{\text{B}} T].
  \end{equation}
  With
  \begin{equation}
  \nu = \frac{n}{(n-2)}
  \end{equation}
  the kinetic energy
  \begin{equation}
  \frac{p^2}{2} =  \frac{n}{2} \,k_{\text{B}} T
  \end{equation}
  as required for an $n$ degree of freedom system at temperature $T$.
  Nevertheless, the important aspect of the dynamics for computing configurational averages
  is the invariant measure, not
  the magnitude of the constrained KE.  The quantity $\nu$
  can then be treated as a free parameter, which can
  be used to move the potential saddle closer to energy $\calE=0$
  (cf.\ Section \ref{sec:hamiltonian_nonhamiltonian}).
  In fact, we shall take $\nu =1$ in our calculations.

  For $n \geq 3$ degrees of freedom, and \emph{provided the dynamics is ergodic on the $\calH = 0$
  energy surface}, Hamiltonian dynamics \eqref{sp_6p}
      will therefore yield the canonical measure in configuration space.  Since the aim  is
      to obtain a canonical distribution, the Hamiltonian dynamics
      derived using \eqref{sp_7} is in this respect  equivalent in principle to the original 
      isokinetic  thermostat \cite{Morriss98}.

  For completeness, we note that, in order to treat the $n=2$ DoF case, 
  it is necessary to use a different Hamiltonian $\calK$ (cf.\ refs \cite{Morriss98,Ezra09}), with
  \begin{equation}
      \label{sp_8}
      \calK =  \frac{1}{2}e^{\betab \Phi} \pi^2 - \frac{\nu}{ 2 \betab} \, e^{-\betab \Phi}.
  \end{equation}
  To obtain the correct canonical measure case we must then take ($n \geqslant 2$)
  \begin{equation}
  \betab = \frac{1}{(n-1)k_{\text{B}} T} = \frac{\beta}{(n-1)}
  \end{equation}
  and
  \begin{equation}
  \nu = \frac{n}{(n-1)} \, .
  \end{equation}

\newpage

\section{Model Hamiltonians}
\label{sec:hamiltonians}

In this section we introduce the model Hamiltonians studied in the
remainder of the paper.  We define systems with three and four DoF suitable for
studying the Hamiltonian isokinetic thermostat on the zero energy surface $\calH=0$.

The phase space structure of these systems is
discussed in Section \ref{sec:hamiltonian_nonhamiltonian},
while numerical computations of gap times, reactive volumes and coordinate distributions
obtained using the Hamiltonian isokinetic thermostat are described in Sec.\ 
\ref{sec:numerical_results}.

The systems considered here consist of $n$ uncoupled oscillators: $n-1$ harmonic modes
plus a single bistable mode.  Although the $n$ modes are uncoupled in the physical Hamiltonian
$H(p, q)$, exponentiation of the potential $\Phi(q)$ introduces intermode coupling in the
isokinetic thermostat Hamiltonian $\calH$.  The bistable mode can be interpreted either as
a thermalizing degree of freedom (following Minary, Martyna and Tuckerman (MMT)
\cite{Minary03a,Minary03b}), or as a reactive degree of freedom  associated with an isomerization
process.  Adopting the latter perspective, we can apply concepts and methods recently
developed for understanding reaction dynamics
(in particular, transition state theory) in phase space 
\cite{wwju,Komatsuzaki02,Komatsuzaki05,Jaffe05,Wiesenfeld05,WaalkensSchubertWiggins08,Ezra09a}
to investigate the problem of thermostat dynamics.

\subsection{Double-well potential}

The double well potential for the thermalizing mode is taken to be a temperature-independent quartic
having the following form:
\begin{equation}
\label{mmt_pot}
\chi(y) = \frac{1}{2} \left(y^4 - \alpha y^2 \right)
\end{equation}
(Note that the Minary, Martyna \& Tuckerman version of the double-well potential is
effectively temperature dependent
\cite{Minary03a,Minary03b}.)
The potential \eqref{mmt_pot} has stationary points at $y=0$ (a maximum)
and $y=\pm \sqrt{\alpha/2}$ (minima),
with corresponding values $0$ and $-\alpha^2 /8$.
Expanding $\chi(y)$ about the minima $y=\pm \sqrt{\alpha/2}$
we find that the effective frequency for harmonic motion in the vicinity of the minima is
\begin{equation}
\omegab = \sqrt{ 2 \alpha}.
\end{equation}

\subsection{Model Hamiltonians}

\subsubsection{Three degrees of freedom}

The 3 DoF potential corresponds to a separable bistable $(y)$ plus harmonic bath modes $(x_1, x_2)$:\begin{equation}
\Phi (x_1, x_2 ,y) = \frac{m \omega_1^2 x_1^2}{2} +
\frac{m \omega_2^2 x_2^2}{2} + \frac{1}{2} \left(y^4 - \alpha y^2 \right).
\label{3dof_pot}
\end{equation}
Setting $m=1$ and choosing potential parameters  $\omega_1 = 1$, $\omega_2 = \sqrt{2}$ and $\alpha=2$ we have
\begin{equation}
\Phi (x_1, x_2 ,y) = \frac{x_1^2}{2} +
x_2^2 +  \frac{y^4}{2} - y^2 .
\label{3dof_pot_a}
\end{equation}
The isokinetic thermostat Hamiltonian $\calH$  is then 
\begin{equation}
\label{eq:ham_3dof}
{\cal H}(x_1, x_2, y, \pi_{x_1},\pi_{x_2}, \pi_y) =
\frac{\pi_{x_1}^2}{2} + \frac{\pi_{x_2}^2}{2} + \frac{\pi_y^2}{2}
- \frac{\nu}{2\beta} \exp \left[ -2 \beta \left\{\frac{ x_1^2}{2} + x_2^2
+ \frac{y^4}{2} - y^2 \right\}\right]. 
\end{equation}

The origin is an equilibrium point of saddle-center-center type with
energy ${\cal H}(0, 0, 0, 0, 0, 0) =-\frac{\nu}{2 \beta}$,
and $\nu$ and $\beta$ are parameters that we can vary.

As mentioned above, the value of the parameter $\nu$ determines the
(constant) value of the physical  kinetic energy $p^2/2$.  It also
determines the energy of the saddle point with respect to the zero of energy, $\calH = 0$.
To obtain the correct
canonical invariant density, it is only necessary that $p^2$ be constant; the
parameter $\nu$ is therefore effectively a free parameter in addition to the temperature
$T$.  

For the numerical computations to be discussed below we take $\alpha =2 $, $\nu = 1$.  
Plots of the $x_2 =0$ slice through the physical potential $\Phi$ for the 3 DoF system 
and the corresponding exponentiated potential ($\beta =1$) are shown in Figure \ref{fig:pot_1}.

\subsubsection{Four degrees of freedom}

The 4 DoF potential corresponds to a separable bistable $(y)$ plus
harmonic bath modes $(x_1, x_2, x_3)$:
\begin{equation}
\Phi (x_1, x_2, x_3, y) = \frac{m \omega_1^2 x_1^2}{2} +
\frac{m \omega_2^2 x_2^2}{2} + \frac{m \omega_3^2 x_3^2}{2}
+ \frac{1}{2} \left(y^4 - \alpha y^2 \right).
\label{4dof_pot}
\end{equation}
Setting $m=1$, $\omega_1 = 1$, $\omega_2 = \sqrt{2}$, $\omega_3 = \sqrt{3}$
and $\alpha=2$ we have
\begin{equation}
\Phi (x_1, x_2, x_3, y) = \frac{x_1^2}{2} +
x_2^2 + \frac{3 x_3^2}{2} + \frac{y^4}{2} - y^2 .
\label{4dof_pot_a}
\end{equation}
From eq.\ \eqref{betab_1},  the parameter $\betab = \beta/2$ for $n=4$ DoF, so that
the isokinetic thermostat Hamiltonian $\calH$  is:
\begin{equation}
\label{eq:ham_4dof}
\begin{split}
{\cal H}(x_1, x_2, x_3, y, \pi_{x_1}, \pi_{x_2}, \pi_{x_3}, \pi_y) & =
\frac{\pi_{x_1}^2}{2} + \frac{\pi_{x_2}^2}{2} + \frac{\pi_{x_3}^2}{2} + \frac{\pi_y^2}{2} \\
& - \frac{\nu}{\beta} \exp \left[ -\beta \left\{\frac{x_1^2}{2} + x_2^2 + \frac{3 x_3^2}{2}
+ \frac{y^4}{2} - y^2 \right\}\right].  
\end{split}
\end{equation}

\noindent The origin is an equilibrium point of saddle-center-center-center type with
energy ${\cal H}(0, 0, 0, 0, 0, 0, 0, 0) =-\frac{\nu}{\beta}$,
and $\nu$ and $\beta$ are parameters that we can vary.
We set the parameter $\nu=1$.

In Section \ref{sec:numerical_results} we present numerical results for both the 
4 DoF and 3 DoF systems corresponding to three values of the temperature:
$\beta =1$, $3$, and $5$.

\newpage

\section{Microcanonical Phase space structure:
Hamiltonian and Non-Hamiltonian Isokinetic Thermostat}
\label{sec:hamiltonian_nonhamiltonian}

In this section, we briefly discuss the phase space structure in the vicinity of the
saddle-center equilibrium points for Hamiltonians \eqref{ham_1} and \eqref{sp_7}.

In previous work, we have established the following \cite{Ezra09}:
\begin{enumerate}

\item If the physical Hamiltonian system defined by eq.\ \eqref{ham_1} has an
equilibrium point of saddle-centre-$\ldots$-centre stability type \cite{Wiggins94} at the origin
then the Hamiltonian system defined by an extended Hamiltonian of the form
\eqref{sp_8} corresponding to the isokinetic
thermostat has a equilibrium point at the origin of saddle-centre-$\ldots$-centre stability type.

\item The energy of the equilibrium  is such that on the zero energy surface
of the extended Hamiltonian $\calK$ the phase space structures
present in the physical Hamiltonian also exist in the extended Hamiltonian phase space.

\item The phase space structures on the zero energy surface corresponding to the
Hamiltonian isokinetic thermostat map to phase space structures in the
non-Hamiltonian thermostatted system obtained by transforming the
Hamiltonian equations of motion for evolution $(\pi, q)$ under $\calK$
to equations of motion for non-canonical variables $(p, q)$ under the constraint of zero energy.

\end{enumerate}

Although these results were previously established for the Hamiltonian $\calK$, eq.\ \eqref{sp_8},
they also hold for Hamiltonian $\calH$, eq.\ \eqref{sp_7} as we describe below.

More precisely, we have the following:  assume that the physical Hamiltonian system \eqref{ham_eq}
has an equilibrium point
at $(q,p) = (q^\ast, p^\ast) = (0,0)$.
The energy of this equilibrium point is $H(0,0) = \Phi(0)$.
The stability of the equilibrium point is determined by the
eigenvalues of the derivative of the Hamiltonian vector field (or
Hessian of the Hamiltonian function) evaluated at the equilibrium
point. This is given by the $2n \times 2n$ matrix:
\begin{equation}
\mbox{Hess}_{\mbox{\tiny sys}} = \left(
\begin{array}{cc}
0_{n \times n} & \mbox{id}_{n \times n} \\
-\Phi_{qq}(0) & 0_{n \times n}
\end{array}
\right), \label{Hess_ham}
\end{equation}
\noindent where $0_{n \times n}$ denotes the $n \times n$ matrix
of zeros and  $\mbox{id}_{n \times n}$ denotes the $n \times n$
identity matrix. We require the equilibrium point to be of
saddle-centre-$\ldots$-centre stability type. This means that the
$2n \times 2n$ matrix $\mbox{Hess}_{\mbox{\tiny sys}}$ has
eigenvalues $\pm \lambda, \pm i \omega_i$, $i=2, \ldots, n$ where
$\lambda$ and $\omega_i$ are real.

Eigenvalues $\gamma$ of $\mbox{Hess}_{\mbox{\tiny sys}}$ are obtained by solving
the characteristic equation
$\det(\mbox{Hess}_{\mbox{\tiny sys}}-\gamma\mbox{id}_{2n \times 2n})=0$.
The block structure of the
$2n \times 2n$ matrix $\mbox{Hess}_{\mbox{\tiny sys}}$ implies that
(cf.\ Theorem 3 of \cite{silvester})
\begin{equation}
\det(\mbox{Hess}_{\mbox{\tiny sys}}  - \gamma \mbox{id}_{2n \times 2n})
= \det(\Phi_{qq}(0) + \gamma^2 \mbox{id}_{n \times n})=0
\label{char_sys}
\end{equation}
\noindent
so that the $2n$ eigenvalues $\gamma$ are given in terms of $\sigma$, the eigenvalues
of the $n \times n$ Hessian matrix $\Phi_{qq} (0)$ associated with the potential, as follows
\begin{equation}
\gamma_{k}, \gamma_{k+n} = \pm \sqrt{-\sigma_k}, \;\;  k=1,\ldots, n.
\label{sys_eivalues}
\end{equation}
\noindent
Therefore, if $\Phi(q)$ has a rank-one saddle at $q=0$, so that one eigenvalue is strictly negative and
the rest are strictly positive, then $(q,p) = (0,0)$ is a
saddle-centre-$\ldots$-centre  type equilibrium point for \eqref{ham_eq} as described above.

Next, we consider Hamilton's equations associated with the extended Hamiltonian \eqref{sp_7},
eq.\ \eqref{sp_6p}, corresponding to the isokinetic thermostat.
It is easy to verify that $(q, \pi) = (0, 0)$ is an
equilibrium point for \eqref{sp_6p} with energy ${\cal H}
(0,0) = -\frac{\nu}{2 \betab} e^{-2 \betab\Phi(0)}$.

Proceeding as above, we  linearize \eqref{sp_6p} about $(q, \pi) = (0, 0)$ 
and compute the eigenvalues of the matrix associated with the linearization. These are given by: 
\begin{equation}
\tilde{\gamma}_{k}, \tilde{\gamma}_{k+n} = \pm \sqrt{\nu} e^{-  \bar{\beta} \Phi (0)}\sqrt{-\sigma_k}, \;\;  k=1,\ldots, n.
\label{therm_eivalues}
\end{equation}
\noindent
In other words, the eigenvalues of the linearization of \eqref{sp_6p} about $(q, \pi) = (0, 0)$ 
correspond to the eigenvalues of  the matrix associated with the 
linearization of \eqref{ham_eq} about $(q, p) = (0, 0)$, but with each 
eigenvalue multiplied by the positive constant  $\sqrt{\nu} e^{-  \bar{\beta} \Phi (0)}$.  
Hence, it follows that
if  the potential of the physical Hamiltonian, $\Phi(q)$,
has a rank-one saddle at $q=0$, so that one eigenvalue is strictly negative and
the rest are strictly positive, then $(q, \pi) = (0,0)$ is a
saddle-centre-$\ldots$-centre  type equilibrium point for \eqref{sp_6p}. 
Moreover, if the purely imaginary eigenvalues in \eqref{sys_eivalues}  satisfy 
a non-resonance condition, then the purely imaginary eigenvalues in \eqref{therm_eivalues} 
satisfy the {\em same} non-resonance condition \cite{Ezra09}.

The equilibrium point $(q,\pi)=(0,0)$ has energy
${\cal H} (0,0) = -\frac{\nu}{2 \betab} e^{-2 \betab \Phi(0)}$.
However, we are only interested in the dynamics on
the ${\cal H}=0$ energy surface. All of the
phase space structure discussed above exists for a certain range of energies above that of the
saddle-centre-$\ldots$-centre, and will do so on the ${\cal H} = 0$ surface if
$-\frac{\nu}{2\betab} e^{-2 \betab\Phi(0)}$ is close enough to zero.

Putting together the results above, we have the following \cite{Ezra09}:
\begin{quotation}
\noindent \em Suppose the physical Hamiltonian system
\eqref{ham_eq} has an equilibrium point of
saddle-centre-$\ldots$-centre stability type at the origin.
Then the Hamiltonian system \eqref{sp_6p}
corresponding to the isokinetic thermostat for $n \geq 3$ DoF  also has an equilibrium point of
saddle-centre-$\ldots$-centre stability type at the origin. Moreover, the energy of this
equilibrium point can be chosen so that on the energy surface corresponding to ${\cal H}=0$
there exists a normally hyperbolic invariant manifold (NHIM) \cite{Wiggins94},
with associated stable and
unstable manifolds,  a ``dividing surface'' of no-return and
minimal flux, and a foliation of the reaction region by $n$-dimensional invariant Lagrangian submanifolds
\cite{WaalkensSchubertWiggins08}.
\end{quotation}

Following the general arguments outlined in \cite{Ezra09},
it is straightforward to show that the phase space structure of
(\ref{sp_6p}) on ${\cal H}=0$ exists in the non-Hamiltonian
thermostatted system in the original physical variables.
These general results allow us to conclude that, under
the noncanonical transformation of variables $(\pi, q) \mapsto (p, q)$,
\begin{itemize}

\item {\em The $2n-1$ dimensional invariant energy surface $\calH =0$ 
corresponding to the Hamiltonian isokinetic thermostat (\ref{sp_6p}) maps to a
$2n-1$ dimensional invariant  manifold for the non-Hamiltonian thermostatted system.}

\item {\em  The $2n-3$ dimensional NHIM, its
$2n-2$ dimensional stable and unstable manifolds, the $n$-dimensional
invariant Lagrangian submanifolds, and the $2n-2$ dimensional dividing surface map
to a $2n-3$ dimensional NHIM, its $2n-2$ dimensional stable and unstable manifolds,
$n$-dimensional invariant submanifolds, and the $2n-2$ dimensional dividing surface
in the $2n-1$ dimensional invariant  manifold for the non-Hamiltonian thermostatted
system.}

\end{itemize}

\newpage

\section{Phase Space Geometrical Structures, Unimolecular Reaction Rates
and Thermostat Dynamics:  Gap Times and Reactive Volumes}
\label{sec:unimolecular}

Our analysis of thermostat dynamics  will be carried out in phase space, and our
approach to probing the dynamics, and especially the question of ergodicity, will be based on the
general formulation of unimolecular reaction
rates based upon that originally given by Thiele \cite{Thiele62}.
In their general form the rate expressions derived by Thiele explicitly
invoke the existence of a \emph{phase space} dividing surface separating
reactants and products; such surfaces, discussed by Wigner \cite{Wigner39}
(see also refs \onlinecite{Keck67,Anderson95}), have only recently become amenable
to direct computation via the use of normal form
approaches \cite{wwju,ujpyw,WaalkensBurbanksWiggins04,WaalkensWiggins04,WaalkensBurbanksWigginsb04,
WaalkensBurbanksWiggins05,WaalkensBurbanksWiggins05c,SchubertWaalkensWiggins06,WaalkensSchubertWiggins08,
Komatsuzaki00,Komatsuzaki02,Komatsuzaki05}.
Our discussion of  Thiele's reaction rate theory
theory follows ref.\ \onlinecite{Ezra09a}.

\subsection{Phase space dividing surfaces:  definition and properties}

Interpreting the bistable thermalizing mode \cite{Minary03a,Minary03b} in the model Hamiltonians of
Section \ref{sec:hamiltonians} as an isomerization coordinate, we see that
the thermostat dynamics associated with Hamiltonian \eqref{sp_7} is equivalent
to an isomerization reaction at constant energy ${\cal H}=0$.
We therefore consider the rate of unimolecular isomerization, at fixed  energy, for a
system described by a time-independent,  $n$ degree-of-freedom (DOF) classical Hamiltonian.

Points in the $2n$-dimensional system phase space $\calM  = \mathbb{R}^{2n}$ are denoted
$\bz \equiv (\pi, q) \in \calM$.
The system Hamiltonian is $\calH(\bz)$, and
the $(2n-1)$ dimensional  energy shell at energy $E$, $\calH(\bz) = E$, is denoted $\Sigma_E \subset \calM$.
The corresponding microcanonical phase
space density is $\delta(E - \calH(\bz))$, and the associated density of states
for the complete energy shell at energy $E$ is
\begin{equation}
\rho(E) = \Int{\bz}{\calM}{} \delta(E - \calH(\bz)).
\end{equation}

In Appendix \ref{sec:dos}, we provide analytical results for $\rho(E)$ for
$n$-dimensional systems with Hamiltonians of the form \eqref{sp_7}, with $\Phi (q)$
an isotropic harmonic potential and $n$ even.

The first step in the analysis is to define regions of the energy surface corresponding to
reactant and product. For the isokinetic thermostat
Hamiltonians considered here, there is a natural divison of
phase space into reactant region ($y<0$, say) and product region
$y >0$.  The dividing surface between reactant and product is determined
by symmetry to be the codimension-1 surface $y=0$, and we shall be concerned with the evaluation of
the microcanonical reactive flux across this surface.
(For fundamental work on reactive flux correlation functions and
associated relaxation kinetics, see \cite{Dumont89,Dumont89a,Chandler78,Chandler87,Gray87}.)

For multidimensional systems such as polyatomic molecules ($n \geq 3$ DoF), it is
\emph{in general} not possible to define or compute a dividing surface with desirable dynamical attributes
such as the no-recrossing property by
working in configuration space alone, and a phase space perspective is necessary
\cite{wwju,ujpyw,WaalkensBurbanksWiggins04,WaalkensWiggins04,WaalkensBurbanksWigginsb04,
WaalkensBurbanksWiggins05,WaalkensBurbanksWiggins05c,SchubertWaalkensWiggins06,WaalkensSchubertWiggins08,
Komatsuzaki00,Komatsuzaki02,Komatsuzaki05}.

As discussed in Section \ref{sec:hamiltonian_nonhamiltonian}, Hamilton's equations for
the thermostat Hamiltonian $\calH$ have an equilibrium of saddle-center-$\ldots$-center stability type.
The significance of saddle points of this type for Hamilton's equations
is that, for a range of energies above that of the saddle,
the energy surfaces have the {\em bottleneck property} in a phase space
neighborhood near the saddle, i.e., the $2n-1$ dimensional
energy surface {\em locally} has the geometrical structure of the product
of a $2n-2$ dimensional sphere and an interval, $S^{2n-2} \times I$.
In the vicinity of the bottleneck,
we are able to construct a dividing surface depending on $E$, $\text{DS}(E)$,
with very desirable properties: 
For each energy in this range above the saddle, $\text{DS}(E)$
locally ``disconnects'' the energy surface into two disjoint pieces
with the consequence that the only way to pass from one
piece of the energy surface to the other is to cross $\text{DS}(E)$.
The dividing surface has the geometrical structure of a
$2n-2$ dimensional sphere, $S^{2n-2}$, which is divided into two $2n-2$ dimensional hemispheres,
denoted $\text{DS}_{\text{in}}(E)$ and $\text{DS}_{\text{out}}(E)$ that are joined at an equator,
which is a $2n-3$ dimensional sphere, $S^{2n-3}$. The hemisphere
$\text{DS}_{\text{in}}(E)$ corresponds to initial conditions of
trajectories that enter the reaction region while $\text{DS}_{\text{out}}(E)$ corresponds to
initial conditions of trajectories that exit the reaction region,
both by passing through the bottleneck in the energy surface.
The equator $S^{2n-3}$ is an invariant manifold of saddle stability type,
a so-called  {\em normally hyperbolic invariant manifold} (NHIM) \cite{Wiggins94}.
The NHIM is of great physical significance: it is the actual ``saddle'' in phase space identified
as the ``activated complex'' of reaction rate dynamics \cite{Pollak78,Truhlar96,WaalkensSchubertWiggins08}.

In the context of microcanonical rates, it
has been shown that $\text{DS}_{\text{in}}(E)$ and $\text{DS}_{\text{out}}(E)$
have the essential \emph{no-recrossing property} and that the flux across them
is minimal \cite{WaalkensWiggins04}. We denote the directional flux
across these hemispheres by $\phi_{\text{in}} (E)$ and $\phi_{\text{out}} (E)$,
respectively, and note that $\phi_{\text{in}} (E)+ \phi_{\text{out}}(E)=0$.
The magnitude of the flux is
$\vert \phi_{\text{in}} (E)\vert =  \vert \phi_{\text{out}}(E)\vert \equiv \phi (E)$.
Most significantly,  the hemisphere  $\text{DS}_{\text{in/out}}(E)$ is the correct surface
across which to compute the ``exact'' flux into/out of the reaction region.

\subsection{Phase space volumes and gap times}
\label{subsec:volumes}

The disjoint regions of phase space corresponding to species A (reactant)
and B (product) will be denoted
$\calM_{\text{A}}$ and $\calM_{\text{B}}$, respectively \cite{footnote0}.

As discussed above, the DS can be rigorously defined to be locally a surface of
no return (transition state).
The microcanonical density of states for reactant species A is
\begin{equation}
\rho_{\text{A}}(E) = \Int{\bz}{\calM_{\text{A}}}{} \delta(E - H(\bz))
\end{equation}
with a corresponding expression for the density of states $\rho_{\text{B}}(E)$
for product B for the case of compact product energy shell.

For isokinetic thermostat Hamiltonians $\calH$, the $\calH = 0$ energy surface
extends to $\pm \infty$ in configuration space.  Although the phase space volume
$N(E)$ is finite for $E\to 0$, the corresponding density of states
$\rho(E) = \rmd N/\rmd E$ may diverge as $E\to 0$.  In Appendix \ref{sec:dos}
we derive analytical expressions for $\rho(E)$ associated with isotropic harmonic 
potentials $\Phi(\bq)$, $\bq \in \mathbb{R}^n$, with $n$  even.  
While $\rho(E)$ diverges as $E \to 0$ for $n=2$,
it is finite in this limit for $n \geq 4$.  Analytical expressions for
$\rho(E)$ are not available for the model potentials studied here.

Provided that the flow is everywhere transverse to $\text{DS}_{\text{in, out}}(E)$,
those phase points  in the reactant region $\calM_{\text{A}}$
that lie on crossing trajectories \cite{DeLeon81,Berne82}
(i.e., that will eventually react)
can be  specified uniquely by coordinates $(\bqb, \bpib, \psi)$,
where $(\bqb, \bpib) \in \text{DS}_{\text{in}}(E)$ is a point on
$\text{DS}_{\text{in}}(E)$, the incoming half of the DS, specified by $2(n-1)$
coordinates $(\bqb, \bpib)$, and
$\psi$ is a time variable.  (Dividing surfaces constructed by
normal form algorithms are guaranteed to be transverse to the vector field, except at the NHIM, where
the vector field is tangent \cite{wwju,ujpyw}.)
The point $\bz(\bqb, \bpib, \psi)$ is reached by propagating the
initial condition $(\bqb, \bpb) \in \text{DS}_{\text{in}}(E)$ forward for time $\psi$
\cite{Thiele62,Ezra09a}.
As all initial conditions on $\text{DS}_{\text{in}}(E)$
(apart from a set of trajectories of measure zero lying on stable manifolds)
will leave the reactant region in finite time by crossing $\text{DS}_{\text{out}}(E)$, for each
$(\bqb, \bpib) \in \text{DS}_{\text{in}}(E)$ we can define the \emph{gap time}
$s = s(\bqb, \bpib)$, which is the
time it takes for the incoming trajectory to traverse the reactant region.
That is, $\bz(\bqb, \bpib, \psi = s(\bqb, \bpb)) \in \text{DS}_{\text{out}}(E)$.
For the phase point $\bz(\bqb, \bpib, \psi)$, we therefore have
$0 \leq \psi \leq s(\bqb, \bpib)$.

The coordinate transformation $\bz \to (E, \psi, \bqb, \bpib)$ is canonical
\cite{Arnold78,Thiele62,Binney85,Meyer86}, so that the phase space volume element is
\begin{equation}
\label{coord_1}
\rmd^{2n} \bz = \rmd E \, \rmd \psi  \, \rmd \sigma
\end{equation}
with $\rmd \sigma \equiv \rmd^{n-1} \qb \, \rmd^{n-1} \bar{\pi}$
an element of $2n-2$ dimensional area on the DS.

The magnitude $\phi(E)$ of the flux through dividing surface
$\text{DS}(E)$ at energy $E$ is given by
\begin{equation}
\label{flux_1}
\phi(E) = \left\vert\Int{\sigma}{\text{DS}_{\text{in}}(E)}{} \right\vert,
\end{equation}
where the element of area $\rmd \sigma$ is precisely the restriction to the DS of the
appropriate flux $(2n-2)$-form $\omega^{(n-1)}/(n-1)!$ corresponding to the Hamiltonian vector field
associated with $H(\bz)$ \cite{Toller85,Mackay90,Gillilan90,WaalkensWiggins04}.
The reactant phase space volume occupied by points initiated on the dividing surface
$\text{DS}_{\text{in}}$ with energies between $E$ and $E + \rmd E$ is therefore
\cite{Thiele62,Brumer80,Pollak81,Binney85,Meyer86,WaalkensBurbanksWiggins05,WaalkensBurbanksWiggins05c}
\begin{subequations}
\label{vol_1}
\begin{align}
\rmd E \Int{\sigma}{\text{DS}_{\text{in}}(E)}{} \Int{\psi}{0}{s}
& = \rmd E \Int{\sigma}{\text{DS}_{\text{in}}(E)}{}  s \\
&= \rmd E \,\, \phi(E) \, \sbar
\end{align}
\end{subequations}where the \emph{mean gap time} $\sbar$ is defined as
\begin{equation}
\sbar = \frac{1}{\phi(E)} \, \Int{\sigma}{\text{DS}_{\text{in}}(E)}{}  s
\end{equation}
and is a function of energy $E$.
The reactant density of states $\rho^{\text{C}}_{\text{A}}(E)$
associated with crossing trajectories only (those trajectories that enter and exit
the reactant region \cite{Berne82}) is then
\begin{equation}
\label{vol_1p}
\rho^{\text{C}}_{\text{A}}(E) = \phi(E) \, \sbar
\end{equation}
where the superscript $\text{C}$ indicates the restriction to crossing trajectories.
The result \eqref{vol_1p} is essentially the content of the so-called
classical spectral theorem
\cite{Brumer80,Pollak81,Binney85,Meyer86,WaalkensBurbanksWiggins05,WaalkensBurbanksWiggins05c}.

If \emph{all} points in the reactant region of phase space eventually react (that is,
all points lie on crossing trajectories
\cite{DeLeon81,Berne82}) then
\begin{equation}
\label{equality_1}
\rho^{\text{C}}_{\text{A}}(E) = \rho_{\text{A}}(E),
\end{equation}
so that the crossing density of states is equal to
the full reactant phase space density of states.
Apart from a set of measure zero, all phase points $\bz \in \calM_{\text{A}}$
can be classified as either trapped (T) or crossing (C) \cite{Berne82}.
A phase point in the trapped region $\calM_{\text{A}}^{\text{T}}$ never crosses the DS,
so that the associated trajectory does not contribute to the reactive flux.
Phase points in the crossing region $\calM_{\text{A}}^{\text{C}}$ do however eventually
cross the dividing surface, and so lie on trajectories that contribute to the reactive flux.
In general, however, as a consequence of the existence of trapped trajectories
(either trajectories on invariant \emph{trapped} $n$-tori \cite{DeLeon81,Berne82} or
trajectories asymptotic to other invariant objects of zero measure),
we have the inequality \cite{Thiele62,Berne82,Hase83}
\begin{equation}
\label{vol_2}
\rho_{\text{A}}^{\text{C}}(E) \leq \rho_{\text{A}}(E).
\end{equation}

From the perspective of thermostat dynamics, the equality \eqref{equality_1}
is a \emph{necessary condition for ergodicity}. If equality \eqref{equality_1}  does not hold, then
there is a region of phase space of nonzero measure that is trapped on the reactant side
of the dividing surface.

If $\rho_{\text{A}}^{\text{C}}(E) < \rho_{\text{A}}(E)$, then it is in principle necessary to
introduce corrections to statistical estimates of reaction rates
\cite{Berne82,Hase83,Gray87,Berblinger94,Grebenshchikov03,Stember07}.
Numerical results for $\rho^{\text{C}}(E)$ and $\rho(E)$ for the HCN molecule are discussed
in \cite{WaalkensBurbanksWigginsb04,WaalkensBurbanksWiggins05,Ezra09a}.

\subsection{Gap time and reactant lifetime distributions}
\label{subsec:gaps}

The \emph{gap time distribution}, $\calP(s; E)$ is of central interest in
unimolecular kinetics \cite{Slater56,Thiele62}: the probability
that a phase point on $\text{DS}_{\text{in}}(E)$ at energy $E$ has a gap time between
$s$ and $s +\rmd s$ is equal to $\calP(s; E) \rmd s$.
An important idealized gap distribution is the random, exponential distribution
\begin{equation}
\label{exp_1}
\calP(s; E) = k(E) \, e^{-k(E) s}
\end{equation}
characterized by a single decay constant $k$ (where $k$ depends on energy $E$),
with corresponding mean gap time $\sbar = k^{-1}$.
An exponential distribution of gap times is  taken to be 
a necessary condition for `statistical' behavior
in unimolecular reactions \cite{Slater56,Slater59,Thiele62,Dumont86}.

The lifetime (time to cross the dividing surface $\text{DS}_{\text{out}}(E)$)
of phase point $\bz(\bqb, \bpb, \psi)$ is $t = s(\bqb, \bpb) - \psi$, and
the corresponding  (normalized)
reactant lifetime distribution function $\bbP(t; E)$ at energy $E$ is 
\cite{Slater56,Slater59,Thiele62,Bunker62,Bunker64,Bunker73,Dumont86}
\begin{subequations}
\label{life_1}
\begin{align}
\label{life_1a}
\bbP(t; E) &= -\frac{\rmd}{\rmd t'}\; \text{Prob}(t \geq t'; E) \Big\vert_{t'=t} \\
\label{life_1b}
&= \frac{1}{\sbar} \, \Int{s}{t}{+\infty} \calP(s; E)
\end{align}
\end{subequations}
where the fraction of interesting (reactive) phase points having lifetimes between $t$ and $t + \rmd t$ is
$\bbP(t; E) \rmd t$.

Equation \eqref{life_1a} gives the general relation between the lifetime distribution and the
fraction of trajectories having lifetimes greater than a certain value for arbitrary ensembles
\cite{Bunker62,Bunker64,Bunker73}.
Note that an exponential gap distribution \eqref{exp_1}
implies that the reactant lifetime
distribution $\bbP(t; E)$ is also exponential 
\cite{Slater56,Slater59,Thiele62,Bunker62,Bunker64,Bunker73}; both gap and lifetime distributions
for realistic molecular potentials have
been of great interest since the earliest days of trajectory simulations of
unimolecular decay, and many examples of non-exponential lifetime distributions
have been found
\cite{Thiele62a,Bunker62,Bunker64,Bunker66,Bunker68,Bunker73,Hase76,Grebenshchikov03,Lourderaj09}.

\subsection{Reaction rates and the inverse gap time}

The quantity
\begin{equation}
\label{k_RRKM}
k^{\text{RRKM}}_f(E) \equiv \frac{\phi(E)}{\rho_{\text{A}}(E)}
\end{equation}
is the statistical (RRKM) microcanonical rate for the forward reaction
(A $\to$ B) at energy $E$, the ratio of the magnitude of
the flux $\phi(E)$ through $\text{DS}_{\text{in}}(E)$
to the total reactant density of states \cite{Robinson72,Forst03}.

Clearly, if $\rho_{\text{A}}(E) = \rho_{\text{A}}^{\text{C}}(E)$, then
\begin{equation}
k^{\text{RRKM}}_f(E) = \frac{1}{\sbar}
\end{equation}
the inverse mean gap time.
In general, the inverse of the mean gap time is
\begin{subequations}
\label{k2}
\begin{align}
\frac{1}{\sbar} &= \frac{\phi(E)}{\rho_{\text{A}}^{\text{C}}} \\ 
& = k^{\text{RRKM}}_f \, \left[\frac{\rho_{\text{A}}(E)}{\rho_{\text{A}}^{\text{C}}(E)}\right] \\
& \geq k^{\text{RRKM}}_f.
\end{align}
\end{subequations}
The inverse gap time can then be interpreted as the
statistical unimolecular reaction  rate corrected for the volume of trapped trajectories in the
reactant phase space \cite{Dumont86,Berne82,Hase83,Gray87,Berblinger94}.

In the next section we discuss our numerical calculations of the following quantities
for the isokinetic thermostats Hamiltonian at constant energy $\calH = 0$:
gap time distributions, mean gap time $\sbar$,
reactive flux $\phi (E=0)$,  reactive volume $\sbar \phi$ and
reactant density of states, $\rho(E=0)$.

\newpage

\section{Numerical computations for isokinetic thermostat}
\label{sec:numerical_results}

\subsection{Computations}

In this Section we discuss the computation of various dynamical
quantities for the isokinetic thermostat Hamiltonians defined in Section
\ref{sec:hamiltonians}.
The potentials for these isokinetic thermostat
Hamiltonians with 3 and 4 DoF 
have the form of an exponentiated double well plus
harmonic modes potential.

As already noted, for the cases examined here, we can exploit the symmetry of the
potential functions. Thus, the dividing surface (DS) between `reactant' and `product'
is simply defined to be the symmetry plane $y=0$ for the double well potential.
For more general non-symmetric potentials, the dividing surface in phase space
can be computed using a normal form expansion \cite{WaalkensSchubertWiggins08}.

\subsubsection{System parameters}

In addition to the frequencies characterizing the modes transverse to the reaction coordinate,
it is necesssary to choose values for the parameters $\beta \equiv 1/\kB T$, $\alpha$, and $\nu$.

We adopt the following notation for presentation of our results:  the system denoted by H$\beta \alpha \nu$
has 3 DoF and parameter values indicated, while the system J$\beta \alpha \nu$ has 4 DoF.
We present numerical results for the 3 DoF systems H121, H321, H521  (Hamiltonian eq.\ \eqref{eq:ham_3dof})
and 4 DoF systems J121, J321, J521 (Hamiltonian eq.\ \eqref{eq:ham_4dof}).

In the units used here the height of the barrier to isomerization in the physical potential
corresponds to  $\beta = 2$.  The temperature values studied here therefore span a 
range of energy scales from well below the barrier height ($\beta = 5$) to well above ($\beta = 1$).

\subsubsection{Computations}

After choosing a set of parameters,
we compute the following quantities (further details of computational 
methodology are given in Appendix \ref{sec:sampling}):

\begin{enumerate}

\item {Gap time distribution and mean gap time}

The distribution of gap times is obtained by starting trajectories on the
forward dividing surface  and propagating them until they cross the
backward dividing surface.  Initial conditions are obtained by uniform random 
sampling of the dividing surface (see Appendix \ref{sec:sampling}).
A discretized approximation to the gap time distribution, $\calP(s)$, is obtained by binning
the gap times for the trajectory ensemble. 
The mean gap time $\sbar$ is calculated as an unweighted average
of computed gap times for the trajectory ensemble.

\item {Lifetime distribution}

The lifetime distribution $\bbP (t)$ is obtained from the (discretized)  gap time distribution 
by numerical integration (cf.\ eq.\ \eqref{life_1}).
The form of the lifetime distibution is of interest: in particular, deviations from exponentiality
are suggestive of ``nonstatistical'' dynamics.

The average lifetime $\langle t \rangle $ for the normalized lifetime distribution
$\bbP(\tau)$ is defined as
\begin{equation}
\langle t \rangle = \Int{t}{0}{\infty} \bbP(t) \, t.
\end{equation}


The random (exponential) lifetime
distribution $\bbP = \bar{k} e^{-\bar{k} t}$ extremizes the information
entropy
\begin{equation}
S_{\bbP} \equiv -\Int{t}{0}{\infty} \bbP(t) \log[\bbP(t)]
\end{equation}
where $\bar{k} = \langle t \rangle^{-1}$ (see, for example, \cite{Chekmarev08}).
One measure of the extent to which a calculated lifetime distribution $\bbP(t)$ 
characterized by mean lifetime $\langle t \rangle$ deviates from exponentiality
is the entropy deficit \cite{Chekmarev08}
\begin{subequations}
\label{delta_s}
\begin{align}
\Delta S_{\bbP} & \equiv S_{\text{Random}} - S(\bbP) \\
& = 1 + \log\langle t \rangle + \Int{t}{0}{\infty} \bbP(t) \log[\bbP(t)].
\end{align}
\end{subequations}


\item {Flux through DS}

The flux through the DS is obtained by integrating the flux form
$\rmd \sigma$ (cf.\ eq.\ \eqref{flux_1}) over the DS.  As the flux form is simply the phase space
volume element associated with the `activated complex', the 
reactive flux $\phi$ can be computed by uniform (random) sampling of the DS phase space 
(see Appendix \ref{sec:sampling}).

The associated reactive volume $2 \phi \times \sbar$ 
is the total phase space volume on the energy shell
$\calH = 0$ traced out by all the trajectories
passing through the dividing surface (in both directions, hence the factor of 2).
Except for a set of measure zero,
each trajectory returns to the DS, and only contributes to the reactive volume up until the gap time.

\item  {Reactant density of states}

The classical density of states $\rho(E)$ of the reactant region at energy $\calH =0$
is obtained by calculating a discretized approximation to $\rho(E)$ as a function of energy 
for $E<0$, fitting $\rho(E)$ to a polynomial function in $E$, and 
evaluating the fitted $\rho(E)$ at $E=0$.
The discretized approximation to 
$\rho(E)$ is obtained by randomly sampling phase points inside a suitably chosen
hypercube, and binning the energies for sampled points with $E \leq 0$.
(See Appendix \ref{sec:sampling}.)

Although in all cases the phase space volume (integrated density of states) $N(E)$ is 
found to be finite as $E \to 0$, for the 3 DoF isokinetic Hamiltonian 
(eq.\ \eqref{eq:ham_3dof}) it is not
clear from our numerical results whether or not the density of states
actually diverges as $E \to 0$, and we have not been able to 
evaluate this limit analytically.
For the isokinetic thermostat with $n=4$ DoF (eq.\ \eqref{eq:ham_4dof}), 
our numerical results suggest that $\rho(E)$ is finite as $E\to 0$.

In Appendix \ref{sec:dos}, we show analytically for 
isotropic harmonic potentials that, 
while $\rho(E)$ diverges logarithmically as $E\to 0$ for $n=2$ DoF, 
for $n\geq 4$ DoF, with $n$ even, $\rho (E=0)$ is finite.

The value of the density of states $\rho(E)$ at $E=0$ is of some interest, as
equality between the reactive density of states determined as 
the product of the mean gap time with the reactive flux and the total reactant 
density of states $\rho(E=0)$ is a \emph{necessary} condition for ergodicity of
the thermostat dynamics.

\item  Thermostat dynamics

It of course important to examine the effectiveness of the
Hamiltonian isokinetic thermostats defined here as thermostats; that is, to assess  
how well time-averaged coordinate
distributions evaluated over a single thermostat trajectory 
reproduce the Boltzmann distributions associated with the relevant
temperature parameter $\beta$.

We therefore pick an initial condition at the coordinate origin on the DS with random momenta 
and $\calH =0$, and propagate it for
a long time ($t=20000$, many periods of the harmonic oscillator modes).
Trajectories are integrated in Mathematica \cite{Mathematica7} 
using the function {\tt NDSolve} with the {\tt SymplecticPartitionedRungeKutta} 
method and fixed step size, $\Delta t = 0.01$. 
By binning coordinate values over such a long trajectory, we obtain a discretized
probability distribution function that
can be compared with the desired canonical (Boltzmann) coordinate distribution.

Moments of powers $x^k$ and $y^k$ obtained as time averages over the trajectory can
also be compared with  thermal averages at temperature $T$.

\end{enumerate}

\subsection{Numerical results: 3 DoF}

\subsubsection{Lifetime distributions}

Computed lifetime distributions for the 3 DoF thermostats H121, H321 and H521 are
shown in Figure \ref{fig:lifetime_3dof}.  It can be seen qualitatively from these plots of
$\log[\bbP (t)]$ versus $t$ that, following initial
transient decay, the lifetime 
distributions become more exponential as the temperature decreases (i.e., as $\beta$ increases). 
The entropic measure of the deviation defined in eq.\ \eqref{delta_s} 
was computed for (renormalized) lifetime decay curves with transients excluded
(that is, we take $t \geq \langle t \rangle$).  The resulting values of $\Delta S$  are
shown in Table \ref{table:gap_times}; the $\Delta S$ values 
reflect the qualitative observation that the lifetime distributions become more exponential 
as $\beta$ increases.

\subsubsection{Phase space volumes}

Numerical values for the flux, mean gap time and reactant phase space volumes are given in 
Table \ref{table:gap_times}.
The magnitude of the reactive flux $\phi$ decreases with temperature
(as $\beta$ increases), as does the inverse gap time.
Smaller inverse gap times therefore correlate with lifetime distributions that
are more nearly exponential (cf.\ \cite{DeLeon81}).

Note that energy surface volumes  for the 3 DoF systems are not shown in Table 
\ref{table:gap_times}.  The reason for this omission is that, on the basis
of our numerical computations (results not shown here), we cannot 
be sure whether or not $\rho(E)$ diverges for 3 DoF systems as $E \to 0$ 
(cf.\ Appendix \ref{sec:dos}).  As we do not have an analytical proof that the
density of states at $E=0$ is finite, we cannot rule out the possibility of
a divergence.

\subsubsection{Thermostat coordinate distributions}

Coordinate distributions for the 3 DoF Hamiltonian isokinetic thermostat 
are shown in Figures \ref{fig:coord1_3dof},  \ref{fig:coord2_3dof} and \ref{fig:coord3_3dof}.
The histogrammed distributions are obtained  by time-averaging over a single trajectory,
while the solid red curves are the associated canonical (Boltzmann) distributions.
Moments $\langle q^k \rangle$ for $q=x_1$ are shown in Figure \ref{fig:moments1_3dof}.

Qualitatively, it is apparent from these results that the effectiveness
of the Hamiltonian isokinetic thermostat, as judged by the similarity of
trajectory-based and Boltzmann coordinate distributions and moments, 
increases with temperature.  Recall that the lifetime distributions 
become more nearly exponential as temperature decreases.

Despite the possibility of an infinite energy shell 
volume, the Hamiltonian isokinetic thermostat for 3 DoF does in fact 
serve to thermalize the 2 uncoupled harmonic modes quite effectively.
Note, however, that even for the long trajectories considered,  
numerical distributions obtained for the thermalizing ($y$) coordinate are not 
symmetric about $y=0$.


\begin{table}[tpb]
\begin{center}
\begin{tabular}{|c|c|c|c|c|c|} \hline
& Mean gap time & Flux & Reactive volume  & Energy surface volume &  $\Delta S$ \\ \hline\hline
H121
 & {16.57} & {6.978} & {231.28} &  -- & 0.034 \\
 \hline
H321
 & {48.88} & {0.773} & {75.61} &  -- & 0.026 \\
 \hline
H521
 & {130.41} & {0.280} & {72.99} &  -- & 0.019 \\
 \hline 
J121
& {12.69} & {41.490} & {1053.36} & {1053.48} & 0.037  \\
 \hline
J321
& {38.51} & {1.536} & {118.31} & {118.66} & 0.021  \\
 \hline
J521
& {101.60} & {0.334} & {67.87} & {69.47} & 0.010 \\
\hline
\end{tabular}
\caption{\label{table:gap_times} Computed mean gap times, fluxes, reactive phase space volumes, 
energy surface volumes and entropy deficits for lifetime distributions
for 3 DoF and 4 DoF model Hamiltonians.  Details of the
computations are discussed in Appendix \ref{sec:sampling}. }
\end{center}
\end{table}%


\subsection{Numerical results: 4 DoF}

\subsubsection{Lifetime distributions}

Computed lifetime distributions for the 4 DoF thermostats J121, J321 and J521 are
shown in Figure \ref{fig:lifetime_4dof}.  
Values of the lifetime distribution entropy deficit $\Delta S$  are
shown in Table \ref{table:gap_times}. As for the 3 DoF systems, 
the 4 DoF lifetime distributions become more exponential as the temperature decreases.

\subsubsection{Phase space volumes}

Numerical values for the flux, mean gap time and reactant phase space volumes for 4 DoF systems are
given in Table \ref{table:gap_times}.
Both the magnitude of the reactive flux $\phi$ and the inverse gap time decrease with temperature
(as $\beta$ increases), while the lifetime decay curves become more nearly
exponential as $\beta$ increases.

Comparison between reactive volumes and total energy surface volumes shows
that both J121 and J321 systems satisfy (at least, within numerical error) 
the necessary condition for ergodicity, while the J521 system
shows a minor deviation from equality.

\subsubsection{Thermostat coordinate distributions}

Distributions for cordinates $x_1$ and $y$ computed 
for the 4 DoF Hamiltonian isokinetic thermostat 
are shown in Figures \ref{fig:coord1_4dof} and \ref{fig:coord4_4dof},
respectively.
The histogrammed distributions are obtained  by time-averaging over a single trajectory,
while the solid red curves are the associated canonical (Boltzmann) distributions.
Moments $\langle q^k \rangle$ for $q=x_1$ are shown in Figure \ref{fig:moments1_4dof}.

As for 3 DoF, the effectiveness
of the Hamiltonian isokinetic thermostat, as judged by the similarity of
trajectory-based and Boltzmann coordinate distributions and moments for the
harmonic oscillator coordinates, 
increases with temperature.  Recall that the lifetime distributions 
become more nearly exponential as temperature decreases.

\newpage

\section{Summary and conclusions}
\label{sec:summary}

In this paper we have investigated the phase space structure and
dynamics of a Hamiltonian isokinetic thermostat. By design,
ergodic thermostat trajectories at fixed (zero) energy generate
a canonical distribution in configuration space \cite{Morriss98,Litniewski93,Morishita03,Ezra09a}.
The physical potentials studied consist of
a single bistable mode (the thermalizing degree of freedom \cite{Minary03a,Minary03b})
plus transverse harmonic modes.  Although these modes are not
coupled in the physical potential, the potential for the Hamiltonian 
thermostat is obtained by exponentiation of the physical potential, which introduces 
coupling between the modes.

Interpreting the bistable mode as a reaction (isomerization) coordinate, we 
are able to establish connections with the theory of unimolecular reaction rates 
\cite{Slater59,Bunker66,Robinson72,Forst03},
in particular the formulation of isomerization rates in terms
of gap times \cite{Slater56,Thiele62}.  In Thiele's general formulation \cite{Thiele62},
the gap time is the time taken for a reactive trajectory initiated on
a dividing surface in phase space to return to the surface. 
(Such phase space dividing surfaces in multidimensional systems 
have been defined and computed using normal form theory 
\cite{wwju,Komatsuzaki02,Komatsuzaki05,Jaffe05,Wiesenfeld05,WaalkensSchubertWiggins08}.) 
The distribution
of gap times for a microcanonical ensemble initiated on the dividing surface is
of great dynamical significance; an exponential distribution of the lifetimes for
the reactive ensemble is usually taken to be an indicator of `statistical' behavior.
Moreover, comparison of the magnitude of the phase space volume swept out by reactive
trajectories as they pass through the reactant region with the total 
phase space volume (classical density of states) for the reactant region
provides a necessary condition for ergodic dynamics.
If the total density of states is appreciably larger than the reactive volume, the
system cannot be ergodic.

We have computed gap times, associated lifetime distributions, mean gap times, reactive
fluxes, reactive  volumes and total reactant phase space volumes for model systems with 3 and 4 DoF.
The symmetry of the model potentials studied means that in all cases the dividing surface is defined
by a single condition on the thermalizing coordinate,  $y=0$.
The thermostats were studied at three different temperatures.
For 4 DoF, the necessary condition for ergodicity is approximately satisfied 
at all three temperatures. 
For both 3 and 4 DoF systems, nonexponential lifetime distributions
are found at high temperatures ($\beta=1$, 
where the potential barrier to isomerization is $1/2$ in the same units), 
while at low temperatures ($\beta=5$)
the lifetime distribution  is more nearly exponential.  

We have quantified the degree of exponentiality of the
lifetime distribution by computing the information entropy deficit 
with respect to pure exponential decay.
From the standpoint of unimolecular reaction rate theory, the decay becomes more
``statistical'' at lower $T$ (smaller flux).  This finding is in accord with the early observations
of Deleon and Berne \cite{DeLeon81} on isomerization dynamics in a model 2-mode system.

We have examined the efficacy of the Hamiltonian isokinetic thermostat by computing
coordinate distributions averaged over a single long trajectory initiated
at random on the dividing surface. 
For the parameter values used here,
coordinate distributions are more nearly canonical (Boltzmann-like) at lower temperatures.

It remains for future work to establish more quantitative
correlations between dynamical attributes of the Hamiltonian thermostat
and thermostat effectiveness.

\begin{acknowledgments}

PC and SW  acknowledge the support of the  Office of Naval Research  
Grant No.~N00014-01-1-0769.  All three authors 
acknowledge the stimulating environment of the NSF sponsored Institute for
Mathematics and its Applications (IMA) at the University of Minnesota,
where the work reported in this paper was begun.

\end{acknowledgments}

\newpage

\appendix

\section{Phase space volume and classical density of states -- 
analytical results for harmonic potentials}
\label{sec:dos}

An important question arising in numerical investigation
of the dynamics and phase space structure of Hamiltonian
isokinetic thermostats concerns the behavior of the density of states $\rho(E)$
in the limit $E\to 0$.
Specifically, it is important to know whether or not
the density of states diverges in this limit.  It can be
difficult to establish convergence of $\rho(E)$ to a finite value as
$E \to 0$ on the basis of numerical calculations of the phase space volume $N(E)$.

In this Appendix, we evaluate analytically the phase space volume $N(E)$ and
the associated density of states $\rho(E) = \rmd N(E)/\rmd E$ for a Hamiltonian of the
form
\begin{equation}
\label{ho_ham}
\calH =  \frac{1}{2}\tilde\pi^2 - \frac{1}{ 2 } \, e^{- \Phi},
  \end{equation}
with
\begin{equation}
\label{ho_pot}
\Phi(q) = \frac{1}{2} \left[\sum_{j = 1}^{n} q_j^2\right],
\end{equation}
and \emph{even} dimensionality $n$.  Physically, the potential
\eqref{ho_pot} corresponds to $n$ uncoupled degenerate harmonic oscillators; we determine the
phase space volume for the exponentiated harmonic potential
appearing in the Hamiltonian \eqref{ho_ham}.

We start by evaluating the phase space volume $N(E)$
enclosed by the energy shell $\calH = E \leq 0$. The potential \eqref{ho_pot} is spherically symmetric,
and so depends only on $r$, the radial coordinate in $n$-dimensional
configuration space.  Let $\rb = \rb(E)$ be the value of $r$ for which the exponentiated
harmonic potential is equal to $E$,
\begin{equation}
\rb = \sqrt{-2\log[-2 E]}.
\end{equation}
Note that, as $E\to0$ from below, $\rb(E)\to\infty$.
At fixed $r$, $0 \leq r \leq \rb$, the magnitude of the momentum $|\tilde\pi|$ is
\begin{subequations}
\begin{align}
|\tilde\pi| & = \sqrt{2E + e^{-r^2/2}} \\
& = e^{-\rb^2/4}\sqrt{e^{(\rb^2-r^2)/2} - 1}
\end{align}
\end{subequations}
If $\calV(x, n)$ is the volume of a ball of radius $x$ in $n$ dimensions,
\begin{equation}
\calV(x,n) = \frac{\pi^{n/2} x^n}{\Gamma[\tfrac{n}{2} + 1]}\,,
\end{equation}
then the total phase space volume $N(E)$ is given by the integral
\begin{subequations}
\begin{align}
N(E) &= \Int{r}{0}{\rb(E)} \calV(|\tilde\pi|, n) \frac{\rmd \calV(r, n)}{\rmd r} \\
 & = \frac{\pi^{n}}{\Gamma[\tfrac{n}{2} + 1]^2} \,  \calI_n
 \end{align}
 \end{subequations}
 with
 \begin{equation}
 \calI_n \equiv \Int{r}{0}{\rb(E)}
 n r^{n-1} e^{- n \rb^2/4}\left[e^{(\rb^2-r^2)/2} - 1\right]^{n/2}.
 \end{equation}
 For even dimension $n$, this integral can be evaluated explicitly.

 For $n=2$ DoF, the integral $\calI_2$ is
 \begin{equation}
 \calI_2 =
 2 - e^{-\rb^2/2} (2 + \rb^2).
 \end{equation}
 In terms of the energy $E$, we have
 \begin{equation}
 N(E) = 2 \pi^2(1 + E (2 - 2 \log[-2 E])),
 \end{equation}
 and the limiting value of the phase space volume is therefore finite
 \begin{equation}
 \lim_{E\to0}N(E)\vert_{n=2} = 2 \pi^2 .
 \end{equation}
 Despite the fact that, when $N(E)$ is plotted as a function of $E$, 
 the derivative of $N(E)$ appears to be finite at $E=0$,
 the density of states $\rho(E)$ is
 \begin{equation}
 \rho(E) = \deriv{N(E)}{E} = -4\pi^2 \log[- 2 E]
 \end{equation}
 which \emph{diverges} as $E\to 0$.
 This divergence is difficult to identify in a plot of $N(E)$ versus $E$ even when the form
 of $N(E)$ is known, and is also difficult to see in plots of numerically determined $\rho(E)$.

 For $n=4$, 
 \begin{equation}
 \calI_4 =
 e^{-\rb^2} \left(\rb^4+6 \rb^2-16 e^{\frac{\rb^2}{2}}+14\right)+2.
   \end{equation}
   In terms of $E$, we have
   \begin{equation}
   N(E) = \tfrac{1}{2} \pi^4 \left(1 + 16 E + 28 E^2 - 24 E^2 \log[-2 E] + 
   8 E^2 \log[-2 E]^2\right),
   \end{equation}
   and the associated $\rho(E)$ is 
   \begin{equation}
   \label{eq:rho_4d}
   \rho(E) = 
   8 \pi^4 \left(1 + 2 E - 2 E \log[-2 E] + E \log[-2 E]^2\right)
   \end{equation}
   which is {finite} as $E \to 0$.  In fact, we have
   \begin{equation}
   \label{eq:4dho_1}
    \lim_{E\to 0} \rho(E) =  8 \pi^4 \simeq 779.27.
   \end{equation}   
   Figure \ref{fig:rho_plot} shows the ratio of the numerically determined $\rho(E)$ for
   the 4 DoF harmonic potential to the theoretical expression \eqref{eq:rho_4d}, over the
   energy range $-0.1 \leq E \leq 0$.  The numerical $\rho(E)$ is obtained by 
   randomly sampling phase space points inside a suitably chosen hypercube, and 
   binning points according to the value of the Hamiltonian.
   The resulting histogram provides a discretized approximation to $\rho(E)$.
   
   The value of $\rho(E=0)$ for the 4D HO potential obtained using the procedure discussed in
   Appendix \ref{sec:sampling} is $773.43$, which is within 1\% of the exact result 
   \eqref{eq:4dho_1}.

To see the structure of these expressions in the general case for $n=2m$, $m$ integer, 
note that the integral
\begin{equation}
\calI_{2m} \equiv \Int{r}{0}{\rb(E)}
 2m r^{2m-1} e^{- m \rb^2/2}\left[e^{(\rb^2-r^2)/2} - 1\right]^{m}
 \end{equation}
is a sum of terms of the form ($c_j$ constant)
\begin{equation}
\calJ_j = c_j e^{-(m-j)\rb^2/2} \Int{r}{0}{\rb(E)}  r^{2m-1} e^{- j r^2/2}, \;\; j=0,1, \ldots, m
\end{equation}
The only term that contributes to $N(E)$ in the limit $E\to 0$ ($\rb\to \infty$) is that with
$j=m$, so that
\begin{equation}
\lim_{E\to 0} N(E) = \frac{\pi^{2m} 2m}{\Gamma[m + 1]^2}  \Int{r}{0}{\infty} r^{2m-1} e^{-m r^2/2}
= \frac{\pi^{2m}}{\Gamma[m+1]} \left(\frac{2}{m}\right)^m.
\end{equation}
The limiting value of the phase space volume is therefore finite for all positive even integers $n=2m$.

To find the general form for the density of states, $\rho(E)$, it is necessary to evaluate
$\rmd \calI_{2m}/\rmd E$.  Differentiating the integral, noting that the integrand vanishes
for $r = \rb(E)$, and using
\begin{equation}
\deriv{E}{\rb} = \frac{\rb}{2} e^{-\rb^2/2}
\end{equation}
we have
\begin{equation}
\deriv{\calI_{2m}}{E}= \Int{r}{0}{\rb(E)}
 (2m)^2 r^{2m-1} e^{- (m-1) \rb^2/2}\left[e^{(\rb^2-r^2)/2} - 1\right]^{m-1}.
 \end{equation}
 It is clear that $\rho(E)$ will diverge as $E\to 0$ \emph{only} for the case $m=1$; for all other
 positive integer $m$, the derivative $\rmd N(E)/\rmd E$ converges to a finite value
 as $E\to 0$.

 We have not been able to derive corresponding analytical expressions for $\rho(E)$ for the isotropic
 harmonic case for odd dimensions.

 \newpage

\section{Phase space sampling and computation of dynamical quantities for
isokinetic thermostats}
\label{sec:sampling}

In this Appendix we provide additional details of the computations  discussed in Section
\ref{sec:numerical_results}.

The potentials for the isokinetic thermostat
Hamiltonians discussed in Sec.\ \ref{sec:hamiltonians} are obtained by exponentiating
the physical potential, which has the form of a one dimensional double well plus uncoupled 
harmonic modes.
For comparison with the analytical results for the density of states for an exponentiated
isotropic harmonic potential given in Appendix \ref{sec:dos}, 
we have also investigated the corresponding Hamiltonians 
numerically for 2,3 and 4 DoF.

As mentioned in Section \ref{sec:numerical_results},
the symmetry of our Hamiltonians means that 
the dividing surface (DS) between `reactant' and `product'
is simply defined to be the symmetry plane $y=0$ for the double well potential.

The results reported in this paper were obtained using
algorithms coded in Python \cite{Lutz06}; 
calculation efficiency
was not a prime objective and could no doubt be 
improved significantly. Many of these
calculations required large numbers of samples and
were therefore subject to a trade-off between sample size and accuracy.
The size of production runs was estimated on the basis of 
results obtained from shorter trial runs.

At the value of the Hamiltonian ${\cal H}=0$,
the energy surface extends to infinity in coordinate ($\bq$) space; the energy
of the saddle point is negative.
As the exponentiated potential is bounded from below,
momentum components are bounded but coordinate components are
unbounded. For phase space points with
larger coordinate component magnitudes, the relative probability of 
having energy  $E \leq 0$ decreases. 
A phase space sampling region
(``box'') is therefore used with appropriate ranges of
the momentum components and an upper limit on
coordinate components chosen to be large enough that the (small) neglected volume 
associated with points having $E \leq 0$ does
not affect the accuracy of the result to within a chosen tolerance at the
sampling density used. 
The largest values for coordinate
component magnitudes were taken to be either 4.0 or 5.0 (in the units used here).

Phase space points are randomly sampled within the box, and  
the required energy surface volume calculated by estimating the 
fraction of points which satisfy the appropriate energy condition.
Increased accuracy requires both a large number of sample points and a large box; 
there is clearly a trade off between calculation accuracy, box size (hence neglected volume
outside the box), sample size and processing time.
We have aimed for an accuracy of one percent while maintaining as far as
possible a common approach across all the parameter sets studied.
A count was taken of points in the periphery (outer 10\% of coordinates) of
the box to check on the validity of this process.

Details of the computations now follow:

\begin{enumerate}

\item {Mean gap time}

The mean gap time is calculated as an appropriately weighted
average over the dividing surface.
We sample the phase space manifold defined by constraints 
$y = 0$, $p_y = 0$. 
Other coordinate and momentum components $(x, p_x)$ were randomly 
sampled within a chosen box. The value of the Hamiltonian was calculated at each point,
and sample points with energies $E>0$ were rejected.  Points
with $E \leq 0$ were used to seed a trajectory from the dividing surface. The value of the
saddle point momentum $p_y$ was calculated (with $p_y \ge 0$) so that the
total energy $\calH = 0$,
giving a trajectory initial condition on the dividing surface.
Note that the momentum $p_y$ is only calculated after a point on the DS has
been accepted.
The trajectory was then integrated using a 4th-order Runge-Kutta algorithm
until it recrossed the surface $y = 0$.
The variation of the energy along the integrated trajectory was monitored to
check stability of the integration procedure.
The maximum cutoff time for the trajectory was either
2000 or 5000 time units (for different parameter sets).
The gap time for each trajectory 
was recorded and the average calculated 
over $\sim$ 100,000 trajectories.

\item {Flux}

The directional flux $\phi$ is the volume of the dividing surface $y=0$ at
$\calH=0$ (cf.\ eq.\ \eqref{flux_1}).
It is calculated by
uniformly sampling the restricted phase space region as above and again calculating the
value of the Hamiltonian to decide whether
each sampled point is within the dividing surface (that is, has energy
$\leq 0$).
The fraction of points accepted times the volume of the sampling
region yields the volume of the DS for $\calH = 0$.

\item {Reactive volume}

The reactive volume is obtained by calculating (twice) the product
of the mean gap time $\sbar$ and the one-way flux $\phi$.
Since only positive values of the  saddle point momentum are used,
only trajectories on the positive side of the dividing
surface are used to calculate the mean gap time.  For symmetric systems
such as those treated here, an additional factor
of 2 is all that is  required to calculate the
total reactive volume on both sides of the dividing surface.

\item {Energy surface volume and density of states}

We wish to calculate the density of states $\rho(E)$ at energy $E=0$,
where $\rho(E)$ is the derivative with respect to energy of $N(E)$,
the total phase space volume enclosed with the energy shell $\calH = 0$.
In order to do so we determine the volume of a set of energy shells of
thickness $\Delta E$.
This data can then be used to obtain a polynomial fit to $\rho(E)$ 
and hence its value at $E=0$.

A phase space sampling region (box) of full phase space dimensionality 
is used, so that we sample in two more dimensions than the computations
described above.
We sample phase space points $\bz \equiv (\pi, q) \in \calM$ with $\calH(\bz)
\leq 0$, and consider the region of phase space occupied by points
with energies $-k \times \Delta E \leq E \leq -(k-1) \times \Delta E$, with $\Delta E$ typically 
$10^{-5}$ and $k$ a positive integer.
The volume of each of these energy shells is estimated by determining the number of points
inside it as a fraction of those inside the full box; the corresponding
density of states
$\rho(E_k)$ at $E_k = -(k-\tfrac12) \times \Delta E$ 
can then be obtained by dividing the volume of the $k$-th shell by $\Delta E$

The density of states $\rho(E)$ can then be calculated using either of two approaches.
The total volume of a set of $k_{\text{max}}$ such shells can be calculated, and the
resulting energy volume $N(E)$ for $- k_{\text{max}} \times \Delta E \leq E \leq 0$ fitted to, 
for example, a polynomial in $E$.  The derivative of the fitting function
can then be evaluated at $\Delta E = 0$.
An alternative procedure is to fit directly the 
estimated values of $\rho(E)$ to a polynomial in $E$ and 
use the resulting fit to obtain $\rho(E)$ at $E=0$.
Both methods should in principle give the same result. 

In the calculations reported here, $\rho(E)$ was calculated for a set of $2 \times 10^3$ shells with 
$\Delta E = 10^{-5}$, and a fifth-order polynomial 
fit  to $\rho(E)$ calculated over the range $-0.02 \leq E \leq 0$.
Noise in the fitted data (see Figure \ref{fig:rho_fit}) can be reduced by smoothing 
(concatenating several energy intervals $\Delta E$), and we have verified that the 
values of $\rho(E=0)$ we obtain are robust with respect to such smoothing.

\end{enumerate}

\newpage


\begin{thebibliography}{100}
\expandafter\ifx\csname bibnamefont\endcsname\relax
  \def\bibnamefont#1{#1}\fi
\expandafter\ifx\csname bibfnamefont\endcsname\relax
  \def\bibfnamefont#1{#1}\fi
\expandafter\ifx\csname url\endcsname\relax
  \def\url#1{\texttt{#1}}\fi
\expandafter\ifx\csname urlprefix\endcsname\relax\def\urlprefix{URL }\fi
\expandafter\ifx\csname bibinfo\endcsname\relax \def\bibinfo#1#2{#2}\fi
\expandafter\ifx\csname eprint\endcsname\relax \def\eprint#1{#1}\fi

\bibitem{Nose91}
\bibinfo{author}{\bibfnamefont{S.}~\bibnamefont{Nos\'{e}}},
  \bibinfo{journal}{Prog. Theo. Phys. Suppl.} \textbf{\bibinfo{volume}{103}},
  \bibinfo{pages}{1} (\bibinfo{year}{1991}).

\bibitem{Morriss98}
\bibinfo{author}{\bibfnamefont{G.~P.} \bibnamefont{Morriss}} \bibnamefont{and}
  \bibinfo{author}{\bibfnamefont{C.~P.} \bibnamefont{Dettmann}},
  \bibinfo{journal}{CHAOS} \textbf{\bibinfo{volume}{8}}, \bibinfo{pages}{321}
  (\bibinfo{year}{1998}).

\bibitem{Hoover04}
\bibinfo{author}{\bibfnamefont{W.~G.} \bibnamefont{Hoover}},
  \bibinfo{author}{\bibfnamefont{K.}~\bibnamefont{Aoki}},
  \bibinfo{author}{\bibfnamefont{C.~G.} \bibnamefont{Hoover}},
  \bibnamefont{and} \bibinfo{author}{\bibfnamefont{S.~V.~D.}
  \bibnamefont{Groot}}, \bibinfo{journal}{Physica D}
  \textbf{\bibinfo{volume}{187}}, \bibinfo{pages}{253} (\bibinfo{year}{2004}).

\bibitem{Leimkuhler04a}
\bibinfo{author}{\bibfnamefont{B.}~\bibnamefont{Leimkuhler}} \bibnamefont{and}
  \bibinfo{author}{\bibfnamefont{S.}~\bibnamefont{Reich}},
  \emph{\bibinfo{title}{{Simulating Hamiltonian Dynamics}}}
  (\bibinfo{publisher}{Cambridge University Press},
  \bibinfo{address}{Cambridge}, \bibinfo{year}{2004}).

\bibitem{Hunenberger05}
\bibinfo{author}{\bibfnamefont{P.}~\bibnamefont{Hunenberger}},
  \bibinfo{journal}{Adv. Polymer Sci.} \textbf{\bibinfo{volume}{173}},
  \bibinfo{pages}{105} (\bibinfo{year}{2005}).

\bibitem{Bond07}
\bibinfo{author}{\bibfnamefont{S.~D.} \bibnamefont{Bond}} \bibnamefont{and}
  \bibinfo{author}{\bibfnamefont{B.~J.} \bibnamefont{Leimkuhler}},
  \bibinfo{journal}{Acta Numerica} \textbf{\bibinfo{volume}{16}},
  \bibinfo{pages}{1} (\bibinfo{year}{2007}).

\bibitem{Evans90}
\bibinfo{author}{\bibfnamefont{D.~J.} \bibnamefont{Evans}} \bibnamefont{and}
  \bibinfo{author}{\bibfnamefont{G.~P.} \bibnamefont{Morriss}},
  \emph{\bibinfo{title}{{Statistical Mechanics of Nonequilibrium Liquids}}}
  (\bibinfo{publisher}{Academic}, \bibinfo{address}{New York},
  \bibinfo{year}{1990}).

\bibitem{Hoover91}
\bibinfo{author}{\bibfnamefont{W.~G.} \bibnamefont{Hoover}},
  \emph{\bibinfo{title}{{Computational Statistical Mechanics}}}
  (\bibinfo{publisher}{Elsevier}, \bibinfo{address}{New York},
  \bibinfo{year}{1991}).

\bibitem{Mundy00}
\bibinfo{author}{\bibfnamefont{C.~J.} \bibnamefont{Mundy}},
  \bibinfo{author}{\bibfnamefont{S.}~\bibnamefont{Balasubramanian}},
  \bibinfo{author}{\bibfnamefont{K.}~\bibnamefont{Bagchi}},
  \bibinfo{author}{\bibfnamefont{M.~E.} \bibnamefont{Tuckerman}},
  \bibinfo{author}{\bibfnamefont{G.~J.} \bibnamefont{Martyna}},
  \bibnamefont{and} \bibinfo{author}{\bibfnamefont{M.~L.} \bibnamefont{Klein}},
  \emph{\bibinfo{title}{{Nonequilibrium Molecular Dynamics}}}
  (\bibinfo{publisher}{Wiley-VCH}, \bibinfo{address}{New York},
  \bibinfo{year}{2000}), vol.~\bibinfo{volume}{14} of
  \emph{\bibinfo{series}{Reviews in Computational Chemistry}}, pp.
  \bibinfo{pages}{291--397}.

\bibitem{RomeroBastida06}
\bibinfo{author}{\bibfnamefont{M.}~\bibnamefont{Romero-Bastida}}
  \bibnamefont{and} \bibinfo{author}{\bibfnamefont{J.~F.}
  \bibnamefont{Aguilar}}, \bibinfo{journal}{J. Phys. A}
  \textbf{\bibinfo{volume}{39}}, \bibinfo{pages}{11155} (\bibinfo{year}{2006}).

\bibitem{Hoover07a}
\bibinfo{author}{\bibfnamefont{W.~G.} \bibnamefont{Hoover}} \bibnamefont{and}
  \bibinfo{author}{\bibfnamefont{C.~G.} \bibnamefont{Hoover}},
  \bibinfo{journal}{J. Chem. Phys.} \textbf{\bibinfo{volume}{126}},
  \bibinfo{pages}{Art. No. 164113} (\bibinfo{year}{2007}).

\bibitem{Jepps10}
\bibinfo{author}{\bibfnamefont{O.~G.} \bibnamefont{Jepps}} \bibnamefont{and}
  \bibinfo{author}{\bibfnamefont{L.}~\bibnamefont{Rondoni}},
  \bibinfo{journal}{J. Phys. A} \textbf{\bibinfo{volume}{43}},
  \bibinfo{pages}{133001} (\bibinfo{year}{2010}).

\bibitem{Nose84}
\bibinfo{author}{\bibfnamefont{S.}~\bibnamefont{Nos\'{e}}},
  \bibinfo{journal}{J. Chem. Phys.} \textbf{\bibinfo{volume}{81}},
  \bibinfo{pages}{511} (\bibinfo{year}{1984}).

\bibitem{Hoover85}
\bibinfo{author}{\bibfnamefont{W.~G.} \bibnamefont{Hoover}},
  \bibinfo{journal}{Phys. Rev. A} \textbf{\bibinfo{volume}{31}},
  \bibinfo{pages}{1695} (\bibinfo{year}{1985}).

\bibitem{Dettmann96}
\bibinfo{author}{\bibfnamefont{C.~P.} \bibnamefont{Dettmann}} \bibnamefont{and}
  \bibinfo{author}{\bibfnamefont{G.~P.} \bibnamefont{Morriss}},
  \bibinfo{journal}{Phys. Rev. E} \textbf{\bibinfo{volume}{54}},
  \bibinfo{pages}{2495} (\bibinfo{year}{1996}).

\bibitem{Dettmann97}
\bibinfo{author}{\bibfnamefont{C.~P.} \bibnamefont{Dettmann}} \bibnamefont{and}
  \bibinfo{author}{\bibfnamefont{G.~P.} \bibnamefont{Morriss}},
  \bibinfo{journal}{Phys. Rev. E} \textbf{\bibinfo{volume}{55}},
  \bibinfo{pages}{3693} (\bibinfo{year}{1997}).

\bibitem{Dettmann99}
\bibinfo{author}{\bibfnamefont{C.~P.} \bibnamefont{Dettmann}},
  \bibinfo{journal}{Phys. Rev. E} \textbf{\bibinfo{volume}{60}},
  \bibinfo{pages}{7576} (\bibinfo{year}{1999}).

\bibitem{Bond99}
\bibinfo{author}{\bibfnamefont{S.~D.} \bibnamefont{Bond}},
  \bibinfo{author}{\bibfnamefont{B.~J.} \bibnamefont{Leimkuhler}},
  \bibnamefont{and} \bibinfo{author}{\bibfnamefont{B.~B.} \bibnamefont{Laird}},
  \bibinfo{journal}{J. Comp. Phys.} \textbf{\bibinfo{volume}{151}},
  \bibinfo{pages}{114} (\bibinfo{year}{1999}).

\bibitem{Tuckerman99}
\bibinfo{author}{\bibfnamefont{M.~E.} \bibnamefont{Tuckerman}},
  \bibinfo{author}{\bibfnamefont{C.~J.} \bibnamefont{Mundy}}, \bibnamefont{and}
  \bibinfo{author}{\bibfnamefont{G.~J.} \bibnamefont{Martyna}},
  \bibinfo{journal}{Europhys. Lett.} \textbf{\bibinfo{volume}{45}},
  \bibinfo{pages}{149} (\bibinfo{year}{1999}).

\bibitem{Tuckerman01}
\bibinfo{author}{\bibfnamefont{M.~E.} \bibnamefont{Tuckerman}},
  \bibinfo{author}{\bibfnamefont{Y.}~\bibnamefont{Liu}},
  \bibinfo{author}{\bibfnamefont{G.}~\bibnamefont{Ciccotti}}, \bibnamefont{and}
  \bibinfo{author}{\bibfnamefont{G.~J.} \bibnamefont{Martyna}},
  \bibinfo{journal}{J. Chem. Phys.} \textbf{\bibinfo{volume}{115}},
  \bibinfo{pages}{1678} (\bibinfo{year}{2001}).

\bibitem{Sergi01}
\bibinfo{author}{\bibfnamefont{A.}~\bibnamefont{Sergi}} \bibnamefont{and}
  \bibinfo{author}{\bibfnamefont{M.}~\bibnamefont{Ferrario}},
  \bibinfo{journal}{Phys. Rev. E} \textbf{\bibinfo{volume}{64}},
  \bibinfo{pages}{Art. No. 056125} (\bibinfo{year}{2001}).

\bibitem{Sergi03}
\bibinfo{author}{\bibfnamefont{A.}~\bibnamefont{Sergi}},
  \bibinfo{journal}{Phys. Rev. E} \textbf{\bibinfo{volume}{67}},
  \bibinfo{pages}{Art. No. 021101} (\bibinfo{year}{2003}).

\bibitem{Ezra04}
\bibinfo{author}{\bibfnamefont{G.~S.} \bibnamefont{Ezra}}, \bibinfo{journal}{J.
  Math. Chem.} \textbf{\bibinfo{volume}{35}}, \bibinfo{pages}{29}
  (\bibinfo{year}{2004}).

\bibitem{Tarasov05}
\bibinfo{author}{\bibfnamefont{V.~E.} \bibnamefont{Tarasov}},
  \bibinfo{journal}{J. Phys. A} \textbf{\bibinfo{volume}{38}},
  \bibinfo{pages}{2145} (\bibinfo{year}{2005}).

\bibitem{Sergi07}
\bibinfo{author}{\bibfnamefont{A.}~\bibnamefont{Sergi}} \bibnamefont{and}
  \bibinfo{author}{\bibfnamefont{P.~V.} \bibnamefont{Giaquinta}},
  \bibinfo{journal}{J. Stat. Mech.} \textbf{\bibinfo{volume}{2007}},
  \bibinfo{pages}{P02013} (\bibinfo{year}{2007}).

\bibitem{Sergi10}
\bibinfo{author}{\bibfnamefont{A.}~\bibnamefont{Sergi}} \bibnamefont{and}
  \bibinfo{author}{\bibfnamefont{G.~S.} \bibnamefont{Ezra}},
  \bibinfo{journal}{Phys. Rev. E} \textbf{\bibinfo{volume}{81}},
  \bibinfo{pages}{036705} (\bibinfo{year}{2010}).

\bibitem{Wojtkowski98}
\bibinfo{author}{\bibfnamefont{M.~P.} \bibnamefont{Wojtkowski}}
  \bibnamefont{and} \bibinfo{author}{\bibfnamefont{C.}~\bibnamefont{Liverani}},
  \bibinfo{journal}{Comm. Math. Phys.} \textbf{\bibinfo{volume}{194}},
  \bibinfo{pages}{47} (\bibinfo{year}{1998}).

\bibitem{Choquard98}
\bibinfo{author}{\bibfnamefont{P.}~\bibnamefont{Choquard}},
  \bibinfo{journal}{CHAOS} \textbf{\bibinfo{volume}{8}}, \bibinfo{pages}{350}
  (\bibinfo{year}{1998}).

\bibitem{Martyna92}
\bibinfo{author}{\bibfnamefont{G.~J.} \bibnamefont{Martyna}},
  \bibinfo{author}{\bibfnamefont{M.~L.} \bibnamefont{Klein}}, \bibnamefont{and}
  \bibinfo{author}{\bibfnamefont{M.~E.} \bibnamefont{Tuckerman}},
  \bibinfo{journal}{J. Chem. Phys.} \textbf{\bibinfo{volume}{97}},
  \bibinfo{pages}{2635} (\bibinfo{year}{1992}).

\bibitem{Kuznezov90}
\bibinfo{author}{\bibfnamefont{D.}~\bibnamefont{Kuznezov}},
  \bibinfo{author}{\bibfnamefont{A.}~\bibnamefont{Bulgac}}, \bibnamefont{and}
  \bibinfo{author}{\bibfnamefont{W.}~\bibnamefont{Bauer}},
  \bibinfo{journal}{Ann. Phys.} \textbf{\bibinfo{volume}{204}},
  \bibinfo{pages}{155} (\bibinfo{year}{1990}).

\bibitem{Bond98}
\bibinfo{author}{\bibfnamefont{S.~D.} \bibnamefont{Bond}} \bibnamefont{and}
  \bibinfo{author}{\bibfnamefont{B.~J.} \bibnamefont{Leimkuhler}},
  \bibinfo{journal}{Num. Algorithms} \textbf{\bibinfo{volume}{19}},
  \bibinfo{pages}{55} (\bibinfo{year}{1998}).

\bibitem{Benest02}
\bibinfo{editor}{\bibfnamefont{D.}~\bibnamefont{Benest}} \bibnamefont{and}
  \bibinfo{editor}{\bibfnamefont{C.}~\bibnamefont{Froeschl\'{e}}}, eds.,
  \emph{\bibinfo{title}{{Singularities in gravitational systems: applications
  to chaotic transport in the solar system}}} (\bibinfo{publisher}{Springer},
  \bibinfo{address}{New York}, \bibinfo{year}{2002}).

\bibitem{SanzSerna94}
\bibinfo{author}{\bibfnamefont{J.~M.} \bibnamefont{Sanz-Serna}}
  \bibnamefont{and} \bibinfo{author}{\bibfnamefont{M.~P.} \bibnamefont{Calvo}},
  \emph{\bibinfo{title}{{Numerical Hamiltonian Problems}}}
  (\bibinfo{publisher}{Chapman and Hall}, \bibinfo{address}{London},
  \bibinfo{year}{1994}).

\bibitem{Hairer02}
\bibinfo{author}{\bibfnamefont{E.}~\bibnamefont{Hairer}},
  \bibinfo{author}{\bibfnamefont{C.}~\bibnamefont{Lubich}}, \bibnamefont{and}
  \bibinfo{author}{\bibfnamefont{G.}~\bibnamefont{Wanner}},
  \emph{\bibinfo{title}{{Geometric Numerical Integration: Structure Preserving
  Algorithms for Ordinary Differential Equations}}}
  (\bibinfo{publisher}{Springer}, \bibinfo{address}{New York},
  \bibinfo{year}{2002}).

\bibitem{Cornfeld82}
\bibinfo{author}{\bibfnamefont{I.~P.} \bibnamefont{Cornfeld}},
  \bibinfo{author}{\bibfnamefont{S.~V.} \bibnamefont{Fomin}}, \bibnamefont{and}
  \bibinfo{author}{\bibfnamefont{Y.~G.} \bibnamefont{Sinai}},
  \emph{\bibinfo{title}{{Ergodic Theory}}} (\bibinfo{publisher}{Springer
  Verlag}, \bibinfo{address}{New York}, \bibinfo{year}{1982}).

\bibitem{Sturman06}
\bibinfo{author}{\bibfnamefont{R.}~\bibnamefont{Sturman}},
  \bibinfo{author}{\bibfnamefont{J.~M.} \bibnamefont{Ottino}},
  \bibnamefont{and} \bibinfo{author}{\bibfnamefont{S.}~\bibnamefont{Wiggins}},
  \emph{\bibinfo{title}{{The Mathematical Foundations of Mixing}}}
  (\bibinfo{publisher}{Cambridge University Press}, \bibinfo{address}{New
  York}, \bibinfo{year}{2006}).

\bibitem{Golo04}
\bibinfo{author}{\bibfnamefont{V.~L.} \bibnamefont{Golo}},
  \bibinfo{author}{\bibfnamefont{V.~N.} \bibnamefont{Salnikov}},
  \bibnamefont{and} \bibinfo{author}{\bibfnamefont{K.~V.}
  \bibnamefont{Shaitan}}, \bibinfo{journal}{Phys. Rev. E}
  \textbf{\bibinfo{volume}{70}}, \bibinfo{pages}{Art. No. 046130}
  (\bibinfo{year}{2004}).

\bibitem{Watanabe07}
\bibinfo{author}{\bibfnamefont{H.}~\bibnamefont{Watanabe}} \bibnamefont{and}
  \bibinfo{author}{\bibfnamefont{H.}~\bibnamefont{Kobayashi}},
  \bibinfo{journal}{Molecular Simulation} \textbf{\bibinfo{volume}{33}},
  \bibinfo{pages}{77} (\bibinfo{year}{2007}).

\bibitem{Watanabe07a}
\bibinfo{author}{\bibfnamefont{H.}~\bibnamefont{Watanabe}} \bibnamefont{and}
  \bibinfo{author}{\bibfnamefont{H.}~\bibnamefont{Kobayashi}},
  \bibinfo{journal}{Phys. Rev. E} \textbf{\bibinfo{volume}{75}},
  \bibinfo{pages}{Art. No. 040102} (\bibinfo{year}{2007}).

\bibitem{Legoll07}
\bibinfo{author}{\bibfnamefont{F.}~\bibnamefont{Legoll}},
  \bibinfo{author}{\bibfnamefont{M.}~\bibnamefont{Luskin}}, \bibnamefont{and}
  \bibinfo{author}{\bibfnamefont{R.}~\bibnamefont{Moeckel}},
  \bibinfo{journal}{Arch. Rat. Mech. Anal.} \textbf{\bibinfo{volume}{184}},
  \bibinfo{pages}{449} (\bibinfo{year}{2007}).

\bibitem{Legoll09}
\bibinfo{author}{\bibfnamefont{F.}~\bibnamefont{Legoll}},
  \bibinfo{author}{\bibfnamefont{M.}~\bibnamefont{Luskin}}, \bibnamefont{and}
  \bibinfo{author}{\bibfnamefont{R.}~\bibnamefont{Moeckel}},
  \bibinfo{journal}{Nonlinearity} \textbf{\bibinfo{volume}{22}},
  \bibinfo{pages}{1673} (\bibinfo{year}{2009}).

\bibitem{Leimkuhler05a}
\bibinfo{author}{\bibfnamefont{B.~J.} \bibnamefont{Leimkuhler}}
  \bibnamefont{and} \bibinfo{author}{\bibfnamefont{C.~R.} \bibnamefont{Sweet}},
  \bibinfo{journal}{SIAM J. Appl. Dyn. Sys.}
  \textbf{\bibinfo{volume}{4}}(\bibinfo{number}{1}), \bibinfo{pages}{187}
  (\bibinfo{year}{2005}).

\bibitem{Tupper05}
\bibinfo{author}{\bibfnamefont{P.~F.} \bibnamefont{Tupper}},
  \bibinfo{journal}{SIAM J. Appl. Dynamical Systems}
  \textbf{\bibinfo{volume}{4}}, \bibinfo{pages}{563} (\bibinfo{year}{2005}).

\bibitem{Skeel09}
\bibinfo{author}{\bibfnamefont{R.~D.} \bibnamefont{Skeel}},
  \bibinfo{journal}{SIAM J. Sci. Comput.} \textbf{\bibinfo{volume}{31}},
  \bibinfo{pages}{1363} (\bibinfo{year}{2009}).

\bibitem{Robinson72}
\bibinfo{author}{\bibfnamefont{P.~J.} \bibnamefont{Robinson}} \bibnamefont{and}
  \bibinfo{author}{\bibfnamefont{K.~A.} \bibnamefont{Holbrook}},
  \emph{\bibinfo{title}{{Unimolecular Reactions}}} (\bibinfo{publisher}{Wiley},
  \bibinfo{address}{New York}, \bibinfo{year}{1972}).

\bibitem{Gilbert90}
\bibinfo{author}{\bibfnamefont{R.~G.} \bibnamefont{Gilbert}} \bibnamefont{and}
  \bibinfo{author}{\bibfnamefont{S.~C.} \bibnamefont{Smith}},
  \emph{\bibinfo{title}{{Theory of Unimolecular and Recombination Reactions}}}
  (\bibinfo{publisher}{Blackwell Scientific}, \bibinfo{address}{Oxford},
  \bibinfo{year}{1990}).

\bibitem{Baer96}
\bibinfo{author}{\bibfnamefont{T.}~\bibnamefont{Baer}} \bibnamefont{and}
  \bibinfo{author}{\bibfnamefont{W.~L.} \bibnamefont{Hase}},
  \emph{\bibinfo{title}{{Unimolecular Reaction Dynamics}}}
  (\bibinfo{publisher}{Oxford University Press}, \bibinfo{address}{New York},
  \bibinfo{year}{1996}).

\bibitem{Forst03}
\bibinfo{author}{\bibfnamefont{W.}~\bibnamefont{Forst}},
  \emph{\bibinfo{title}{{Unimolecular Reactions}}}
  (\bibinfo{publisher}{Cambridge University Press},
  \bibinfo{address}{Cambridge}, \bibinfo{year}{2003}).

\bibitem{Brumer88}
\bibinfo{author}{\bibfnamefont{P.}~\bibnamefont{Brumer}} \bibnamefont{and}
  \bibinfo{author}{\bibfnamefont{M.}~\bibnamefont{Shapiro}},
  \bibinfo{journal}{Adv. Chem. Phys.} \textbf{\bibinfo{volume}{70}},
  \bibinfo{pages}{365} (\bibinfo{year}{1988}).

\bibitem{Rice81}
\bibinfo{author}{\bibfnamefont{S.~A.} \bibnamefont{Rice}},
  \bibinfo{journal}{Adv. Chem. Phys.} \textbf{\bibinfo{volume}{XLVII}},
  \bibinfo{pages}{117} (\bibinfo{year}{1981}).

\bibitem{DeLeon81}
\bibinfo{author}{\bibfnamefont{N.}~\bibnamefont{DeLeon}} \bibnamefont{and}
  \bibinfo{author}{\bibfnamefont{B.~J.} \bibnamefont{Berne}},
  \bibinfo{journal}{J. Chem. Phys.} \textbf{\bibinfo{volume}{75}},
  \bibinfo{pages}{3495} (\bibinfo{year}{1981}).

\bibitem{Rice96}
\bibinfo{author}{\bibfnamefont{S.~A.} \bibnamefont{Rice}} \bibnamefont{and}
  \bibinfo{author}{\bibfnamefont{M.~S.} \bibnamefont{Zhao}},
  \bibinfo{journal}{Int. J. Quantum Chem.} \textbf{\bibinfo{volume}{58}},
  \bibinfo{pages}{593} (\bibinfo{year}{1996}).

\bibitem{Carpenter05}
\bibinfo{author}{\bibfnamefont{B.~K.} \bibnamefont{Carpenter}},
  \bibinfo{journal}{Ann. Rev. Phys. Chem.} \textbf{\bibinfo{volume}{56}},
  \bibinfo{pages}{57} (\bibinfo{year}{2005}).

\bibitem{Ezra09a}
\bibinfo{author}{\bibfnamefont{G.~S.} \bibnamefont{Ezra}},
  \bibinfo{author}{\bibfnamefont{H.}~\bibnamefont{Waalkens}}, \bibnamefont{and}
  \bibinfo{author}{\bibfnamefont{S.}~\bibnamefont{Wiggins}},
  \bibinfo{journal}{J. Chem. Phys.} \textbf{\bibinfo{volume}{130}},
  \bibinfo{pages}{164118} (\bibinfo{year}{2009}).

\bibitem{Wiggins90}
\bibinfo{author}{\bibfnamefont{S.}~\bibnamefont{Wiggins}},
  \bibinfo{journal}{Physica D} \textbf{\bibinfo{volume}{44}},
  \bibinfo{pages}{471} (\bibinfo{year}{1990}).

\bibitem{Wiggins92}
\bibinfo{author}{\bibfnamefont{S.}~\bibnamefont{Wiggins}},
  \emph{\bibinfo{title}{Chaotic Transport in Dynamical Systems}}
  (\bibinfo{publisher}{Springer-Verlag}, \bibinfo{year}{1992}).

\bibitem{Wiggins94}
\bibinfo{author}{\bibfnamefont{S.}~\bibnamefont{Wiggins}},
  \emph{\bibinfo{title}{Normally Hyperbolic Invariant Manifolds in Dynamical
  Systems}} (\bibinfo{publisher}{Springer-Verlag}, \bibinfo{year}{1994}).

\bibitem{wwju}
\bibinfo{author}{\bibfnamefont{S.}~\bibnamefont{Wiggins}},
  \bibinfo{author}{\bibfnamefont{L.}~\bibnamefont{Wiesenfeld}},
  \bibinfo{author}{\bibfnamefont{C.}~\bibnamefont{Jaffe}}, \bibnamefont{and}
  \bibinfo{author}{\bibfnamefont{T.}~\bibnamefont{Uzer}},
  \bibinfo{journal}{Phys. Rev. Lett.} \textbf{\bibinfo{volume}{86(24)}},
  \bibinfo{pages}{5478} (\bibinfo{year}{2001}).

\bibitem{Komatsuzaki02}
\bibinfo{author}{\bibfnamefont{T.}~\bibnamefont{Komatsuzaki}} \bibnamefont{and}
  \bibinfo{author}{\bibfnamefont{R.~S.} \bibnamefont{Berry}},
  \bibinfo{journal}{Adv. Chem. Phys.} \textbf{\bibinfo{volume}{123}},
  \bibinfo{pages}{79} (\bibinfo{year}{2002}).

\bibitem{Komatsuzaki05}
\bibinfo{author}{\bibfnamefont{T.}~\bibnamefont{Komatsuzaki}},
  \bibinfo{author}{\bibfnamefont{K.}~\bibnamefont{Hoshino}}, \bibnamefont{and}
  \bibinfo{author}{\bibfnamefont{Y.}~\bibnamefont{Matsunaga}},
  \bibinfo{journal}{Adv. Chem. Phys.} \textbf{\bibinfo{volume}{130 B}},
  \bibinfo{pages}{257} (\bibinfo{year}{2005}).

\bibitem{Jaffe05}
\bibinfo{author}{\bibfnamefont{C.}~\bibnamefont{Jaff\'{e}}},
  \bibinfo{author}{\bibfnamefont{K.}~\bibnamefont{Shinnosuke}},
  \bibinfo{author}{\bibfnamefont{J.}~\bibnamefont{Palacian}},
  \bibinfo{author}{\bibfnamefont{P.}~\bibnamefont{Yanguas}}, \bibnamefont{and}
  \bibinfo{author}{\bibfnamefont{T.}~\bibnamefont{Uzer}},
  \bibinfo{journal}{Adv. Chem. Phys.} \textbf{\bibinfo{volume}{130 A}},
  \bibinfo{pages}{171} (\bibinfo{year}{2005}).

\bibitem{Wiesenfeld05}
\bibinfo{author}{\bibfnamefont{L.}~\bibnamefont{Wiesenfeld}},
  \bibinfo{journal}{Adv. Chem. Phys.} \textbf{\bibinfo{volume}{130 A}},
  \bibinfo{pages}{217} (\bibinfo{year}{2005}).

\bibitem{WaalkensSchubertWiggins08}
\bibinfo{author}{\bibfnamefont{H.}~\bibnamefont{Waalkens}},
  \bibinfo{author}{\bibfnamefont{R.}~\bibnamefont{Schubert}}, \bibnamefont{and}
  \bibinfo{author}{\bibfnamefont{S.}~\bibnamefont{Wiggins}},
  \bibinfo{journal}{Nonlinearity}
  \textbf{\bibinfo{volume}{21}}(\bibinfo{number}{1}), \bibinfo{pages}{R1}
  (\bibinfo{year}{2008}).

\bibitem{Pechukas81}
\bibinfo{author}{\bibfnamefont{P.}~\bibnamefont{Pechukas}},
  \bibinfo{journal}{Ann. Rev. Phys. Chem.} \textbf{\bibinfo{volume}{32}},
  \bibinfo{pages}{159} (\bibinfo{year}{1981}).

\bibitem{Lourderaj09}
\bibinfo{author}{\bibfnamefont{L.}~\bibnamefont{U.}} \bibnamefont{and}
  \bibinfo{author}{\bibfnamefont{W.~L.} \bibnamefont{Hase}},
  \bibinfo{journal}{J. Phys. Chem. A} \textbf{\bibinfo{volume}{113}},
  \bibinfo{pages}{2236} (\bibinfo{year}{2009}).

\bibitem{Posch86}
\bibinfo{author}{\bibfnamefont{H.~A.} \bibnamefont{Posch}},
  \bibinfo{author}{\bibfnamefont{W.~G.} \bibnamefont{Hoover}},
  \bibnamefont{and} \bibinfo{author}{\bibfnamefont{F.~J.}
  \bibnamefont{Vesely}}, \bibinfo{journal}{Phys. Rev. A}
  \textbf{\bibinfo{volume}{33}}, \bibinfo{pages}{4253} (\bibinfo{year}{1986}).

\bibitem{Posch97}
\bibinfo{author}{\bibfnamefont{H.~A.} \bibnamefont{Posch}} \bibnamefont{and}
  \bibinfo{author}{\bibfnamefont{W.~G.} \bibnamefont{Hoover}},
  \bibinfo{journal}{Phys. Rev. E} \textbf{\bibinfo{volume}{55}},
  \bibinfo{pages}{6803} (\bibinfo{year}{1997}).

\bibitem{Hoover98b}
\bibinfo{author}{\bibfnamefont{W.~G.} \bibnamefont{Hoover}},
  \bibinfo{journal}{J. Chem. Phys.} \textbf{\bibinfo{volume}{109}},
  \bibinfo{pages}{4164} (\bibinfo{year}{1998}).

\bibitem{Hoover01a}
\bibinfo{author}{\bibfnamefont{W.~G.} \bibnamefont{Hoover}},
  \bibinfo{author}{\bibfnamefont{C.~G.} \bibnamefont{Hoover}},
  \bibnamefont{and} \bibinfo{author}{\bibfnamefont{D.~J.}
  \bibnamefont{Isbister}}, \bibinfo{journal}{Phys. Rev. E}
  \textbf{\bibinfo{volume}{63}}, \bibinfo{pages}{026209}
  (\bibinfo{year}{2001}).

\bibitem{DAlessandro02}
\bibinfo{author}{\bibfnamefont{M.}~\bibnamefont{D'Alessandro}},
  \bibinfo{author}{\bibfnamefont{A.}~\bibnamefont{Tenenbaum}},
  \bibnamefont{and} \bibinfo{author}{\bibfnamefont{A.}~\bibnamefont{Amadei}},
  \bibinfo{journal}{J. Phys. Chem. B} \textbf{\bibinfo{volume}{106}},
  \bibinfo{pages}{5050} (\bibinfo{year}{2002}).

\bibitem{Minary03a}
\bibinfo{author}{\bibfnamefont{P.}~\bibnamefont{Minary}},
  \bibinfo{author}{\bibfnamefont{G.~J.} \bibnamefont{Martyna}},
  \bibnamefont{and} \bibinfo{author}{\bibfnamefont{M.~E.}
  \bibnamefont{Tuckerman}}, \bibinfo{journal}{J. Chem. Phys.}
  \textbf{\bibinfo{volume}{118}}, \bibinfo{pages}{2510} (\bibinfo{year}{2003}).

\bibitem{Minary03b}
\bibinfo{author}{\bibfnamefont{P.}~\bibnamefont{Minary}},
  \bibinfo{author}{\bibfnamefont{G.~J.} \bibnamefont{Martyna}},
  \bibnamefont{and} \bibinfo{author}{\bibfnamefont{M.~E.}
  \bibnamefont{Tuckerman}}, \bibinfo{journal}{J. Chem. Phys.}
  \textbf{\bibinfo{volume}{118}}, \bibinfo{pages}{2527} (\bibinfo{year}{2003}).

\bibitem{Litniewski93}
\bibinfo{author}{\bibfnamefont{M.}~\bibnamefont{Litniewski}},
  \bibinfo{journal}{J. Phys. Chem.} \textbf{\bibinfo{volume}{97}},
  \bibinfo{pages}{3842} (\bibinfo{year}{1993}).

\bibitem{Morishita03}
\bibinfo{author}{\bibfnamefont{T.}~\bibnamefont{Morishita}},
  \bibinfo{journal}{J. Chem. Phys.} \textbf{\bibinfo{volume}{119}},
  \bibinfo{pages}{7075} (\bibinfo{year}{2003}).

\bibitem{Ezra09}
\bibinfo{author}{\bibfnamefont{G.~S.} \bibnamefont{Ezra}} \bibnamefont{and}
  \bibinfo{author}{\bibfnamefont{S.}~\bibnamefont{Wiggins}},
  \bibinfo{journal}{J. Phys. A} \textbf{\bibinfo{volume}{42}},
  \bibinfo{pages}{042001} (\bibinfo{year}{2009}).

\bibitem{Slater56}
\bibinfo{author}{\bibfnamefont{N.~B.} \bibnamefont{Slater}},
  \bibinfo{journal}{J. Chem. Phys.}
  \textbf{\bibinfo{volume}{24}}(\bibinfo{number}{6}), \bibinfo{pages}{1256}
  (\bibinfo{year}{1956}).

\bibitem{Thiele62}
\bibinfo{author}{\bibfnamefont{E.}~\bibnamefont{Thiele}}, \bibinfo{journal}{J.
  Chem. Phys.} \textbf{\bibinfo{volume}{36}}(\bibinfo{number}{6}),
  \bibinfo{pages}{1466} (\bibinfo{year}{1962}).

\bibitem{Dumont86}
\bibinfo{author}{\bibfnamefont{R.~S.} \bibnamefont{Dumont}} \bibnamefont{and}
  \bibinfo{author}{\bibfnamefont{P.}~\bibnamefont{Brumer}},
  \bibinfo{journal}{J. Phys. Chem.} \textbf{\bibinfo{volume}{90}},
  \bibinfo{pages}{3509} (\bibinfo{year}{1986}).

\bibitem{Brumer80}
\bibinfo{author}{\bibfnamefont{P.}~\bibnamefont{Brumer}},
  \bibinfo{author}{\bibfnamefont{D.~E.} \bibnamefont{Fitz}}, \bibnamefont{and}
  \bibinfo{author}{\bibfnamefont{D.}~\bibnamefont{Wardlaw}},
  \bibinfo{journal}{J. Chem. Phys.}
  \textbf{\bibinfo{volume}{72}}(\bibinfo{number}{1}), \bibinfo{pages}{386}
  (\bibinfo{year}{1980}).

\bibitem{Pollak81}
\bibinfo{author}{\bibfnamefont{E.}~\bibnamefont{Pollak}}, \bibinfo{journal}{J.
  Chem. Phys.} \textbf{\bibinfo{volume}{74}}, \bibinfo{pages}{6763}
  (\bibinfo{year}{1981}).

\bibitem{Binney85}
\bibinfo{author}{\bibfnamefont{J.}~\bibnamefont{Binney}},
  \bibinfo{author}{\bibfnamefont{O.~E.} \bibnamefont{Gerhard}},
  \bibnamefont{and} \bibinfo{author}{\bibfnamefont{P.}~\bibnamefont{Hut}},
  \bibinfo{journal}{Mon. Not. Roy. Astron. Soc.}
  \textbf{\bibinfo{volume}{215}}, \bibinfo{pages}{59} (\bibinfo{year}{1985}).

\bibitem{Meyer86}
\bibinfo{author}{\bibfnamefont{H.-D.} \bibnamefont{Meyer}},
  \bibinfo{journal}{J. Chem. Phys.} \textbf{\bibinfo{volume}{84}},
  \bibinfo{pages}{3147} (\bibinfo{year}{1986}).

\bibitem{WaalkensBurbanksWiggins05}
\bibinfo{author}{\bibfnamefont{H.}~\bibnamefont{Waalkens}},
  \bibinfo{author}{\bibfnamefont{A.}~\bibnamefont{Burbanks}}, \bibnamefont{and}
  \bibinfo{author}{\bibfnamefont{S.}~\bibnamefont{Wiggins}},
  \bibinfo{journal}{Physical Review Letters} \textbf{\bibinfo{volume}{95}},
  \bibinfo{pages}{084301} (\bibinfo{year}{2005}).

\bibitem{WaalkensBurbanksWiggins05c}
\bibinfo{author}{\bibfnamefont{H.}~\bibnamefont{Waalkens}},
  \bibinfo{author}{\bibfnamefont{A.}~\bibnamefont{Burbanks}}, \bibnamefont{and}
  \bibinfo{author}{\bibfnamefont{S.}~\bibnamefont{Wiggins}},
  \bibinfo{journal}{J. Phys. A} \textbf{\bibinfo{volume}{38}},
  \bibinfo{pages}{L759} (\bibinfo{year}{2005}).

\bibitem{Evans83}
\bibinfo{author}{\bibfnamefont{D.~J.} \bibnamefont{Evans}},
  \bibinfo{author}{\bibfnamefont{W.~G.} \bibnamefont{Hoover}},
  \bibinfo{author}{\bibfnamefont{B.~H.} \bibnamefont{Failor}},
  \bibinfo{author}{\bibfnamefont{B.}~\bibnamefont{Moran}}, \bibnamefont{and}
  \bibinfo{author}{\bibfnamefont{A.~J.~C.} \bibnamefont{Ladd}},
  \bibinfo{journal}{Phys. Rev. A} \textbf{\bibinfo{volume}{28}},
  \bibinfo{pages}{1016} (\bibinfo{year}{1983}).

\bibitem{Evans83a}
\bibinfo{author}{\bibfnamefont{D.~J.} \bibnamefont{Evans}} \bibnamefont{and}
  \bibinfo{author}{\bibfnamefont{G.~P.} \bibnamefont{Morriss}},
  \bibinfo{journal}{Phys. Lett. A} \textbf{\bibinfo{volume}{98}},
  \bibinfo{pages}{433} (\bibinfo{year}{1983}).

\bibitem{Szebehely67}
\bibinfo{author}{\bibfnamefont{V.}~\bibnamefont{Szebehely}},
  \emph{\bibinfo{title}{{Theory of orbits}}} (\bibinfo{publisher}{Academic
  Press}, \bibinfo{address}{New York}, \bibinfo{year}{1967}).

\bibitem{Verlet67}
\bibinfo{author}{\bibfnamefont{L.}~\bibnamefont{Verlet}},
  \bibinfo{journal}{Phys. Rev.} \textbf{\bibinfo{volume}{159}},
  \bibinfo{pages}{98} (\bibinfo{year}{1967}).

\bibitem{silvester}
\bibinfo{author}{\bibfnamefont{J.~R.} \bibnamefont{Silvester}},
  \bibinfo{journal}{The Mathematical Gazette}
  \textbf{\bibinfo{volume}{84}}(\bibinfo{number}{501}), \bibinfo{pages}{460}
  (\bibinfo{year}{2000}).

\bibitem{Wigner39}
\bibinfo{author}{\bibfnamefont{E.~P.} \bibnamefont{Wigner}},
  \bibinfo{journal}{J. Chem. Phys.} \textbf{\bibinfo{volume}{7}},
  \bibinfo{pages}{646} (\bibinfo{year}{1939}).

\bibitem{Keck67}
\bibinfo{author}{\bibfnamefont{J.~C.} \bibnamefont{Keck}},
  \bibinfo{journal}{Adv. Chem. Phys.} \textbf{\bibinfo{volume}{XIII}},
  \bibinfo{pages}{85} (\bibinfo{year}{1967}).

\bibitem{Anderson95}
\bibinfo{author}{\bibfnamefont{J.~B.} \bibnamefont{Anderson}},
  \bibinfo{journal}{Adv. Chem. Phys.} \textbf{\bibinfo{volume}{XCI}},
  \bibinfo{pages}{381} (\bibinfo{year}{1995}).

\bibitem{ujpyw}
\bibinfo{author}{\bibfnamefont{T.}~\bibnamefont{Uzer}},
  \bibinfo{author}{\bibfnamefont{C.}~\bibnamefont{Jaffe}},
  \bibinfo{author}{\bibfnamefont{J.}~\bibnamefont{Palacian}},
  \bibinfo{author}{\bibfnamefont{P.}~\bibnamefont{Yanguas}}, \bibnamefont{and}
  \bibinfo{author}{\bibfnamefont{S.}~\bibnamefont{Wiggins}},
  \bibinfo{journal}{Nonlinearity} \textbf{\bibinfo{volume}{15}},
  \bibinfo{pages}{957} (\bibinfo{year}{2002}).

\bibitem{WaalkensBurbanksWiggins04}
\bibinfo{author}{\bibfnamefont{H.}~\bibnamefont{Waalkens}},
  \bibinfo{author}{\bibfnamefont{A.}~\bibnamefont{Burbanks}}, \bibnamefont{and}
  \bibinfo{author}{\bibfnamefont{S.}~\bibnamefont{Wiggins}},
  \bibinfo{journal}{J. Phys. A} \textbf{\bibinfo{volume}{37}},
  \bibinfo{pages}{L257} (\bibinfo{year}{2004}).

\bibitem{WaalkensWiggins04}
\bibinfo{author}{\bibfnamefont{H.}~\bibnamefont{Waalkens}} \bibnamefont{and}
  \bibinfo{author}{\bibfnamefont{S.}~\bibnamefont{Wiggins}},
  \bibinfo{journal}{J. Phys. A} \textbf{\bibinfo{volume}{37}},
  \bibinfo{pages}{L435} (\bibinfo{year}{2004}).

\bibitem{WaalkensBurbanksWigginsb04}
\bibinfo{author}{\bibfnamefont{H.}~\bibnamefont{Waalkens}},
  \bibinfo{author}{\bibfnamefont{A.}~\bibnamefont{Burbanks}}, \bibnamefont{and}
  \bibinfo{author}{\bibfnamefont{S.}~\bibnamefont{Wiggins}},
  \bibinfo{journal}{J. Chem. Phys.}
  \textbf{\bibinfo{volume}{121}}(\bibinfo{number}{13}), \bibinfo{pages}{6207}
  (\bibinfo{year}{2004}).

\bibitem{SchubertWaalkensWiggins06}
\bibinfo{author}{\bibfnamefont{R.}~\bibnamefont{Schubert}},
  \bibinfo{author}{\bibfnamefont{H.}~\bibnamefont{Waalkens}}, \bibnamefont{and}
  \bibinfo{author}{\bibfnamefont{S.}~\bibnamefont{Wiggins}},
  \bibinfo{journal}{Phys. Rev. Lett.} \textbf{\bibinfo{volume}{96}},
  \bibinfo{pages}{218302} (\bibinfo{year}{2006}).

\bibitem{Komatsuzaki00}
\bibinfo{author}{\bibfnamefont{T.}~\bibnamefont{Komatsuzaki}} \bibnamefont{and}
  \bibinfo{author}{\bibfnamefont{R.~S.} \bibnamefont{Berry}},
  \bibinfo{journal}{J. Mol. Struct. THEOCHEM} \textbf{\bibinfo{volume}{506}},
  \bibinfo{pages}{55} (\bibinfo{year}{2000}).

\bibitem{Dumont89}
\bibinfo{author}{\bibfnamefont{R.~S.} \bibnamefont{Dumont}},
  \bibinfo{journal}{J. Chem. Phys.} \textbf{\bibinfo{volume}{91}},
  \bibinfo{pages}{4679} (\bibinfo{year}{1989}).

\bibitem{Dumont89a}
\bibinfo{author}{\bibfnamefont{R.~S.} \bibnamefont{Dumont}},
  \bibinfo{journal}{J. Chem. Phys.} \textbf{\bibinfo{volume}{91}},
  \bibinfo{pages}{6839} (\bibinfo{year}{1989}).

\bibitem{Chandler78}
\bibinfo{author}{\bibfnamefont{D.}~\bibnamefont{Chandler}},
  \bibinfo{journal}{J. Chem. Phys.} \textbf{\bibinfo{volume}{68}},
  \bibinfo{pages}{2959} (\bibinfo{year}{1978}).

\bibitem{Chandler87}
\bibinfo{author}{\bibfnamefont{D.}~\bibnamefont{Chandler}},
  \emph{\bibinfo{title}{{Introduction to Modern Statistical Mechanics}}}
  (\bibinfo{publisher}{Oxford University Press}, \bibinfo{address}{New York},
  \bibinfo{year}{1987}).

\bibitem{Gray87}
\bibinfo{author}{\bibfnamefont{S.~K.} \bibnamefont{Gray}} \bibnamefont{and}
  \bibinfo{author}{\bibfnamefont{S.~A.} \bibnamefont{Rice}},
  \bibinfo{journal}{J. Chem. Phys.} \textbf{\bibinfo{volume}{86}},
  \bibinfo{pages}{2020} (\bibinfo{year}{1987}).

\bibitem{Pollak78}
\bibinfo{author}{\bibfnamefont{E.}~\bibnamefont{Pollak}} \bibnamefont{and}
  \bibinfo{author}{\bibfnamefont{P.}~\bibnamefont{Pechukas}},
  \bibinfo{journal}{J. Chem. Phys.} \textbf{\bibinfo{volume}{69}},
  \bibinfo{pages}{1218} (\bibinfo{year}{1978}).

\bibitem{Truhlar96}
\bibinfo{author}{\bibfnamefont{D.~G.} \bibnamefont{Truhlar}},
  \bibinfo{author}{\bibfnamefont{B.~C.} \bibnamefont{Garrett}},
  \bibnamefont{and} \bibinfo{author}{\bibfnamefont{S.~J.}
  \bibnamefont{Klippenstein}}, \bibinfo{journal}{J. Phys. Chem.}
  \textbf{\bibinfo{volume}{100}}, \bibinfo{pages}{12771}
  (\bibinfo{year}{1996}).

\bibitem{footnote0}
\bibinfo{note}{{Such a separation is assumed to be meaningful for the range of
  energies considered here.}}

\bibitem{Berne82}
\bibinfo{author}{\bibfnamefont{B.~J.} \bibnamefont{Berne}},
  \bibinfo{author}{\bibfnamefont{N.}~\bibnamefont{DeLeon}}, \bibnamefont{and}
  \bibinfo{author}{\bibfnamefont{R.~O.} \bibnamefont{Rosenberg}},
  \bibinfo{journal}{J. Phys. Chem.} \textbf{\bibinfo{volume}{86}},
  \bibinfo{pages}{2166} (\bibinfo{year}{1982}).

\bibitem{Arnold78}
\bibinfo{author}{\bibfnamefont{V.~I.} \bibnamefont{Arnold}},
  \emph{\bibinfo{title}{{Mathematical Methods of Classical Mechanics}}}
  (\bibinfo{publisher}{Springer-Verlag}, \bibinfo{address}{New York},
  \bibinfo{year}{1978}).

\bibitem{Toller85}
\bibinfo{author}{\bibfnamefont{M.}~\bibnamefont{Toller}},
  \bibinfo{author}{\bibfnamefont{G.}~\bibnamefont{Jacucci}},
  \bibinfo{author}{\bibfnamefont{G.}~\bibnamefont{DeLorenzi}},
  \bibnamefont{and} \bibinfo{author}{\bibfnamefont{C.~P.} \bibnamefont{Flynn}},
  \bibinfo{journal}{Phys. Rev. B} \textbf{\bibinfo{volume}{32}},
  \bibinfo{pages}{2082} (\bibinfo{year}{1985}).

\bibitem{Mackay90}
\bibinfo{author}{\bibfnamefont{R.~S.} \bibnamefont{MacKay}},
  \bibinfo{journal}{Phys. Lett. A} \textbf{\bibinfo{volume}{145}},
  \bibinfo{pages}{425} (\bibinfo{year}{1990}).

\bibitem{Gillilan90}
\bibinfo{author}{\bibfnamefont{R.~E.} \bibnamefont{Gillilan}},
  \bibinfo{journal}{J. Chem. Phys.} \textbf{\bibinfo{volume}{93}},
  \bibinfo{pages}{5300} (\bibinfo{year}{1990}).

\bibitem{Hase83}
\bibinfo{author}{\bibfnamefont{W.~L.} \bibnamefont{Hase}},
  \bibinfo{author}{\bibfnamefont{D.~G.} \bibnamefont{Buckowski}},
  \bibnamefont{and} \bibinfo{author}{\bibfnamefont{K.~N.} \bibnamefont{Swamy}},
  \bibinfo{journal}{J. Phys. Chem.} \textbf{\bibinfo{volume}{87}},
  \bibinfo{pages}{2754} (\bibinfo{year}{1983}).

\bibitem{Berblinger94}
\bibinfo{author}{\bibfnamefont{M.}~\bibnamefont{Berblinger}} \bibnamefont{and}
  \bibinfo{author}{\bibfnamefont{C.}~\bibnamefont{Schlier}},
  \bibinfo{journal}{J. Chem. Phys.} \textbf{\bibinfo{volume}{101}},
  \bibinfo{pages}{4750} (\bibinfo{year}{1994}).

\bibitem{Grebenshchikov03}
\bibinfo{author}{\bibfnamefont{S.~Y.} \bibnamefont{Grebenshchikov}},
  \bibinfo{author}{\bibfnamefont{R.}~\bibnamefont{Schinke}}, \bibnamefont{and}
  \bibinfo{author}{\bibfnamefont{W.~L.} \bibnamefont{Hase}}, in
  \emph{\bibinfo{booktitle}{{Unimolecular Kinetics: Part 1. The Reaction
  Step}}}, edited by \bibinfo{editor}{\bibfnamefont{N.~J.~B.}
  \bibnamefont{Greene}} (\bibinfo{publisher}{Elsevier}, \bibinfo{address}{New
  York}, \bibinfo{year}{2003}), vol.~\bibinfo{volume}{{39}} of
  \emph{\bibinfo{series}{{Comprehensive Chemical Kinetics}}}, pp.
  \bibinfo{pages}{105--242}.

\bibitem{Stember07}
\bibinfo{author}{\bibfnamefont{J.~N.} \bibnamefont{Stember}} \bibnamefont{and}
  \bibinfo{author}{\bibfnamefont{G.~S.} \bibnamefont{Ezra}},
  \bibinfo{journal}{Chem. Phys.} \textbf{\bibinfo{volume}{337}},
  \bibinfo{pages}{11} (\bibinfo{year}{2007}).

\bibitem{Slater59}
\bibinfo{author}{\bibfnamefont{N.~B.} \bibnamefont{Slater}},
  \emph{\bibinfo{title}{{Theory of Unimolecular Reactions}}}
  (\bibinfo{publisher}{Cornell University Press}, \bibinfo{address}{Ithaca,
  NY}, \bibinfo{year}{1959}).

\bibitem{Bunker62}
\bibinfo{author}{\bibfnamefont{D.~L.} \bibnamefont{Bunker}},
  \bibinfo{journal}{J. Chem. Phys.} \textbf{\bibinfo{volume}{37}},
  \bibinfo{pages}{393} (\bibinfo{year}{1962}).

\bibitem{Bunker64}
\bibinfo{author}{\bibfnamefont{D.~L.} \bibnamefont{Bunker}},
  \bibinfo{journal}{J. Chem. Phys.} \textbf{\bibinfo{volume}{40}},
  \bibinfo{pages}{1946} (\bibinfo{year}{1964}).

\bibitem{Bunker73}
\bibinfo{author}{\bibfnamefont{D.~L.} \bibnamefont{Bunker}} \bibnamefont{and}
  \bibinfo{author}{\bibfnamefont{W.~L.} \bibnamefont{Hase}},
  \bibinfo{journal}{J. Chem. Phys.} \textbf{\bibinfo{volume}{59}},
  \bibinfo{pages}{4621} (\bibinfo{year}{1973}).

\bibitem{Thiele62a}
\bibinfo{author}{\bibfnamefont{E.}~\bibnamefont{Thiele}}, \bibinfo{journal}{J.
  Chem. Phys.} \textbf{\bibinfo{volume}{38}}(\bibinfo{number}{8}),
  \bibinfo{pages}{1959} (\bibinfo{year}{1963}).

\bibitem{Bunker66}
\bibinfo{author}{\bibfnamefont{D.~L.} \bibnamefont{Bunker}},
  \emph{\bibinfo{title}{{Theory of Elementary Gas Reaction Rates}}}
  (\bibinfo{publisher}{Pergamon}, \bibinfo{address}{Oxford},
  \bibinfo{year}{1966}).

\bibitem{Bunker68}
\bibinfo{author}{\bibfnamefont{D.~L.} \bibnamefont{Bunker}} \bibnamefont{and}
  \bibinfo{author}{\bibfnamefont{M.~L.} \bibnamefont{Pattengil}},
  \bibinfo{journal}{J. Chem. Phys.} \textbf{\bibinfo{volume}{48}},
  \bibinfo{pages}{772} (\bibinfo{year}{1968}).

\bibitem{Hase76}
\bibinfo{author}{\bibfnamefont{W.~L.} \bibnamefont{Hase}}, in
  \emph{\bibinfo{booktitle}{{Modern Theoretical Chemistry}}}, edited by
  \bibinfo{editor}{\bibfnamefont{W.~H.} \bibnamefont{Miller}}
  (\bibinfo{publisher}{Plenum}, \bibinfo{address}{New York},
  \bibinfo{year}{1976}), vol.~\bibinfo{volume}{2}, pp.
  \bibinfo{pages}{121--170}.

\bibitem{Chekmarev08}
\bibinfo{author}{\bibfnamefont{S.~F.} \bibnamefont{Chekmarev}},
  \bibinfo{journal}{Phys. Rev. E} \textbf{\bibinfo{volume}{78}},
  \bibinfo{pages}{Article Number: 066113} (\bibinfo{year}{2008}).

\bibitem{Mathematica7}
\bibinfo{author}{\bibnamefont{{Wolfram Research, Inc.}}},
  \emph{\bibinfo{title}{{Mathematica, V7}}} (\bibinfo{address}{Champaign,
  Illinois}, \bibinfo{year}{2008}).

\bibitem{Lutz06}
\bibinfo{author}{\bibfnamefont{M.}~\bibnamefont{Lutz}},
  \emph{\bibinfo{title}{{Programming Python}}} (\bibinfo{publisher}{O'Reilly
  Media}, \bibinfo{address}{New York}, \bibinfo{year}{2006}), 3rd ed.

\end{thebibliography}

\def\cprime{$'$} \def\cprime{$'$}

\newpage

\section*{Figure captions}

\begin{figure}[H]
 \caption{(a) Section ($x_2 =0$) through the physical potential $\Phi(x_1, x_2, y)$,
 eq.\ \eqref{3dof_pot_a}, $\omega_1 =1$, $\omega_2 = \sqrt{2}$, $\alpha=2$.  
 (b)  Section ($x_2=0$) through the exponentiated potential,
 $-\frac{1}{2}\exp[-2 \beta \Phi]$, $\beta = 1$.} 
 \label{fig:pot_1} 
\end{figure}

\begin{figure}[H]
 \caption{Lifetime distributions for the 3 DoF Hamiltonian isokinetic thermostat.
 The lifetime distribution is derived from the distribution of gap times obtained
 by initiating trajectories on the incoming DS and propagating them until they
 cross the outgoing DS.
 (a)  H121, $\beta = 1$. (b) H321,  $\beta = 3$. (c) H521, $\beta = 5$.
 } 
 \label{fig:lifetime_3dof} 
\end{figure}

\begin{figure}[H]
 \caption{Coordinate distributions for 3 DoF Hamiltonian isokinetic thermostat, obtained
 by averaging over a single trajectory.  Numerical distributions for the
 $x_1$ coordinate (histograms) are compared
 with the Boltzmann distribution (solid line).
 (a)  H121, $\beta = 1$. (b) H321,  $\beta = 3$. (c) H521, $\beta = 5$.} 
 \label{fig:coord1_3dof} 
\end{figure}

\begin{figure}[H]
 \caption{Coordinate distributions for 3 DoF Hamiltonian isokinetic thermostat, obtained
 by averaging over a single trajectory.  Numerical distributions for the
 $x_2$ coordinate (histograms) are compared
 with the Boltzmann distribution (solid line).
 (a)  H121, $\beta = 1$. (b) H321,  $\beta = 3$. (c) H521, $\beta = 5$.} 
 \label{fig:coord2_3dof} 
\end{figure}

\begin{figure}[H]
 \caption{Coordinate distributions for 3 DoF Hamiltonian isokinetic thermostat, obtained
 by averaging over a single trajectory.  Numerical distributions for the
 $y$ coordinate (histograms) are compared
 with the Boltzmann distribution (solid line).
 (a)  H121, $\beta = 1$. (b) H321,  $\beta = 3$. (c) H521, $\beta = 5$.} 
 \label{fig:coord3_3dof} 
\end{figure}

\begin{figure}[H]
 \caption{Moments of the distribution of the coordinate $x_1$ obtained 
 using the 3 DoF Hamiltonian isokinetic thermostat (squares) are compared with
 those for the Boltzmann distribution (circles).  Odd moments for the Boltzmann distribution are
 identically zero. 
 (a)  H121, $\beta = 1$. (b) H321,  $\beta = 3$. (c) H521, $\beta = 5$.} 
 \label{fig:moments1_3dof} 
\end{figure}

\begin{figure}[H]
 \caption{Lifetime distributions for the 4 DoF Hamiltonian isokinetic thermostat.
 The lifetime distribution is derived from the distribution of gap times obtained
 by initiating trajectories on the incoming DS and propagating them until they
 cross the outgoing DS.
 (a)  J121, $\beta = 1$. (b) J321,  $\beta = 3$. (c) J521, $\beta = 5$.
 } 
 \label{fig:lifetime_4dof} 
\end{figure}

\begin{figure}[H]
 \caption{Coordinate distributions for 4 DoF Hamiltonian isokinetic thermostat, obtained
 by averaging over a single trajectory.  Numerical distributions for the
 $x_1$ coordinate (histograms) are compared
 with the Boltzmann distribution (solid line).
 (a)  J121, $\beta = 1$. (b) J321,  $\beta = 3$. (c) J521, $\beta = 5$.} 
 \label{fig:coord1_4dof} 
\end{figure}

\begin{figure}[H]
 \caption{Coordinate distributions for 4 DoF Hamiltonian isokinetic thermostat, obtained
 by averaging over a single trajectory.  Numerical distributions for the
 $y$ coordinate (histograms) are compared
 with the Boltzmann distribution (solid line).
 (a)  J121, $\beta = 1$. (b) J321,  $\beta = 3$. (c) J521, $\beta = 5$.} 
 \label{fig:coord4_4dof} 
\end{figure}

\begin{figure}[H]
 \caption{Moments of the distribution of the coordinate $x_1$ obtained 
 using the 4 DoF Hamiltonian isokinetic thermostat (squares) are compared with
 those for the Boltzmann distribution (circles).  Odd moments for the Boltzmann distribution are
 identically zero. 
 (a)  J121, $\beta = 1$. (b) J321,  $\beta = 3$. (c) J521, $\beta = 5$.} 
 \label{fig:moments1_4dof} 
\end{figure}

\begin{figure}[H]
 \caption{Ratio of the numerically determined $\rho(E)$ for
   the 4 DoF harmonic potential to the theoretical expression \eqref{eq:rho_4d}, over the
   energy range $-0.1 \leq E \leq 0$.  The phase space volume $N(E)$ was determined by random 
   sampling of a phase space hypercube using $N_p = 5\times 10^7$ points.} 
 \label{fig:rho_plot} 
\end{figure}

\begin{figure}[H]
 \caption{Number of points per energy bin versus energy $E$
 (width $\Delta E = 10^{-5}$, $-0.02 \leq E \leq 0$) for 
 the J321 4 DoF Hamiltonian.  The red curve is a fit to the numerical data using a 
 5-th order polynomial.  The fit to the data yields a value $\rho(E=0) = 118.66$.
 } 
 \label{fig:rho_fit} 
\end{figure}



\newpage

\begin{figure}[htb!]
 \begin{center}
 \includegraphics[width=3.5in]{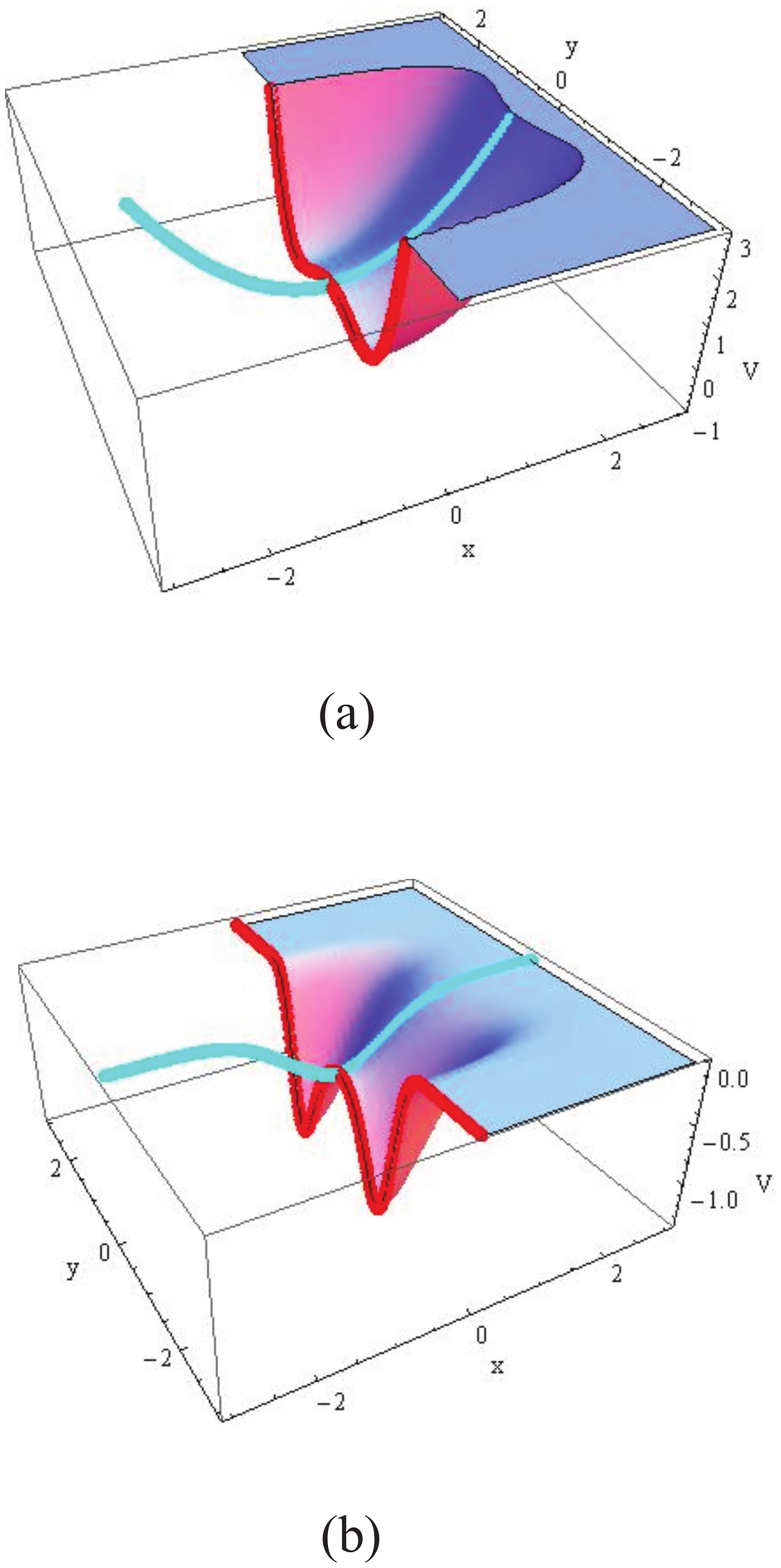}
 \end{center}
\end{figure}

   \vspace*{1.5cm}
   FIGURE 1

\newpage

\begin{figure}[htb!]
 \begin{center}
 \includegraphics[width=3.5in]{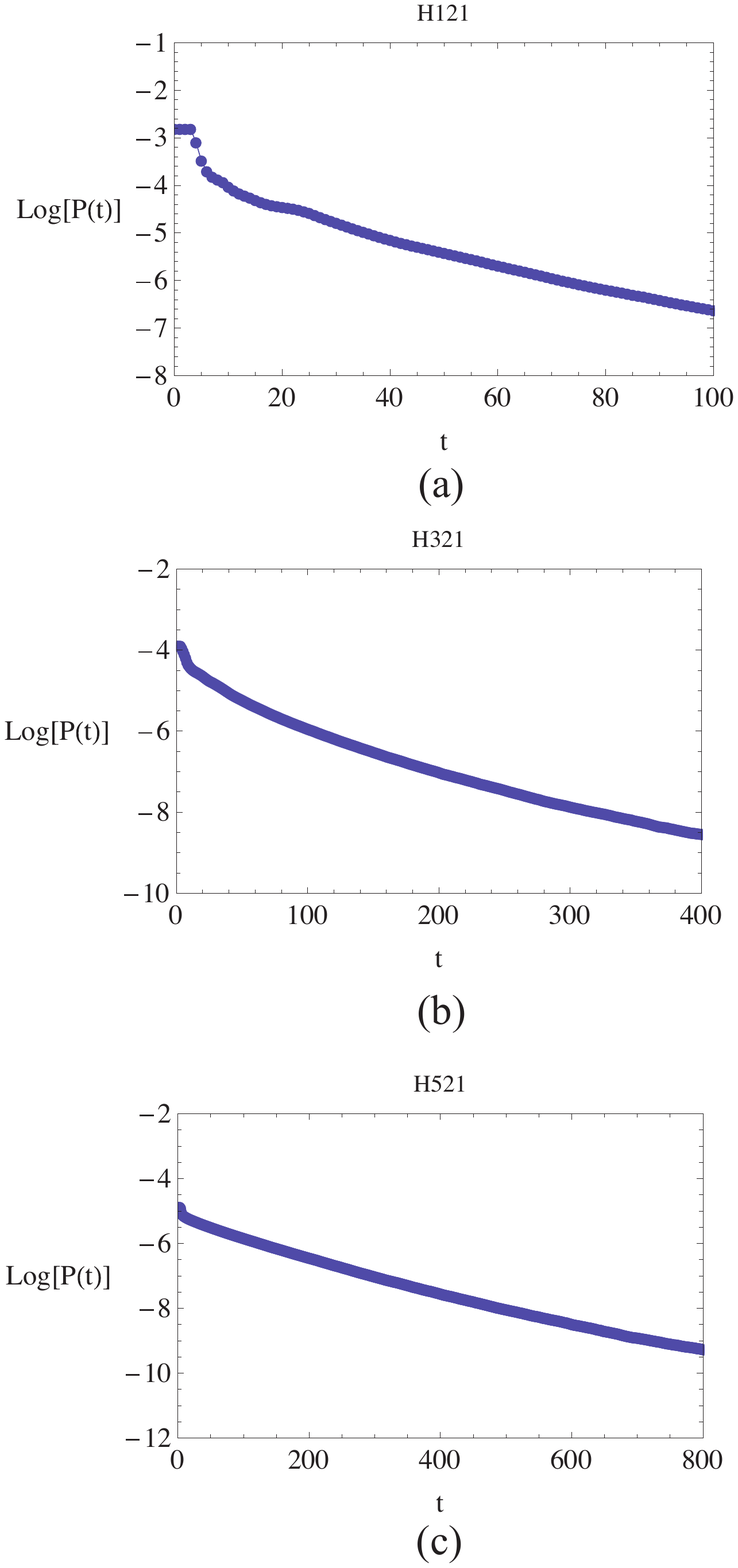}
 \end{center}
\end{figure}

   \vspace*{1.5cm}
   FIGURE 2

\newpage

\begin{figure}[htb!]
 \begin{center}
 \includegraphics[width=3.5in]{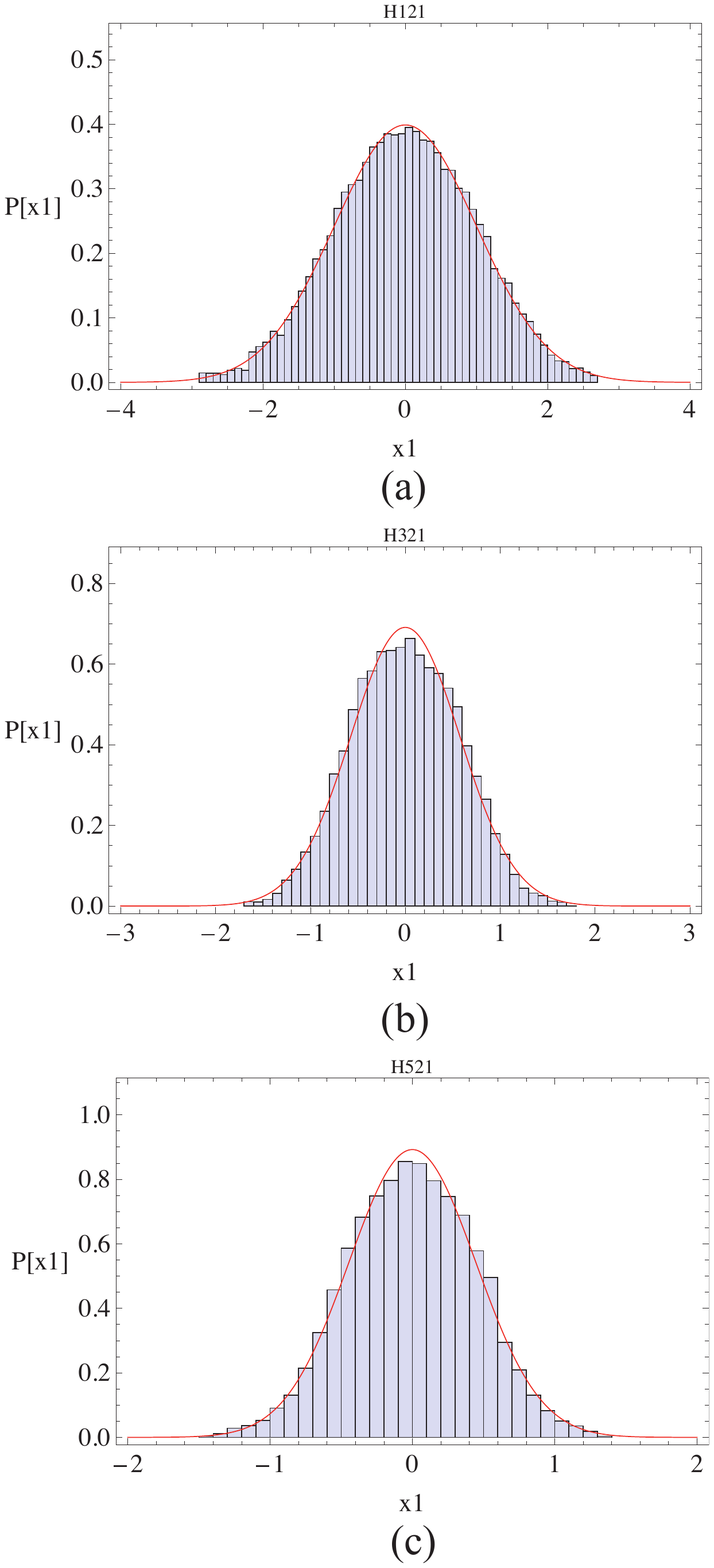}
 \end{center}
 \end{figure}

   \vspace*{1.5cm}
   FIGURE 3

\newpage

\begin{figure}[htb!]
 \begin{center}
 \includegraphics[width=3.5in]{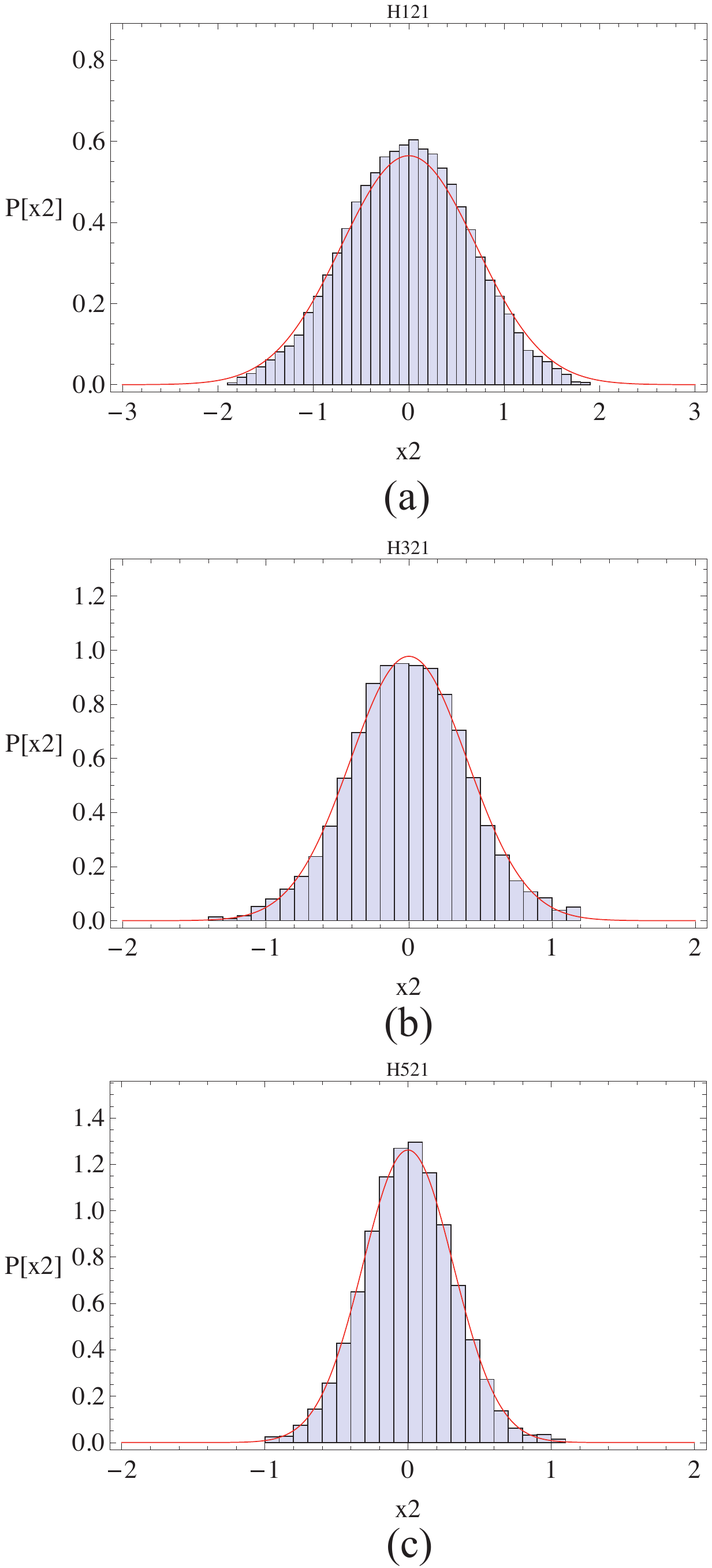}
 \end{center}
 \end{figure}

   \vspace*{1.5cm}
   FIGURE 4

\newpage

\begin{figure}[htb!]
 \begin{center}
 \includegraphics[width=3.5in]{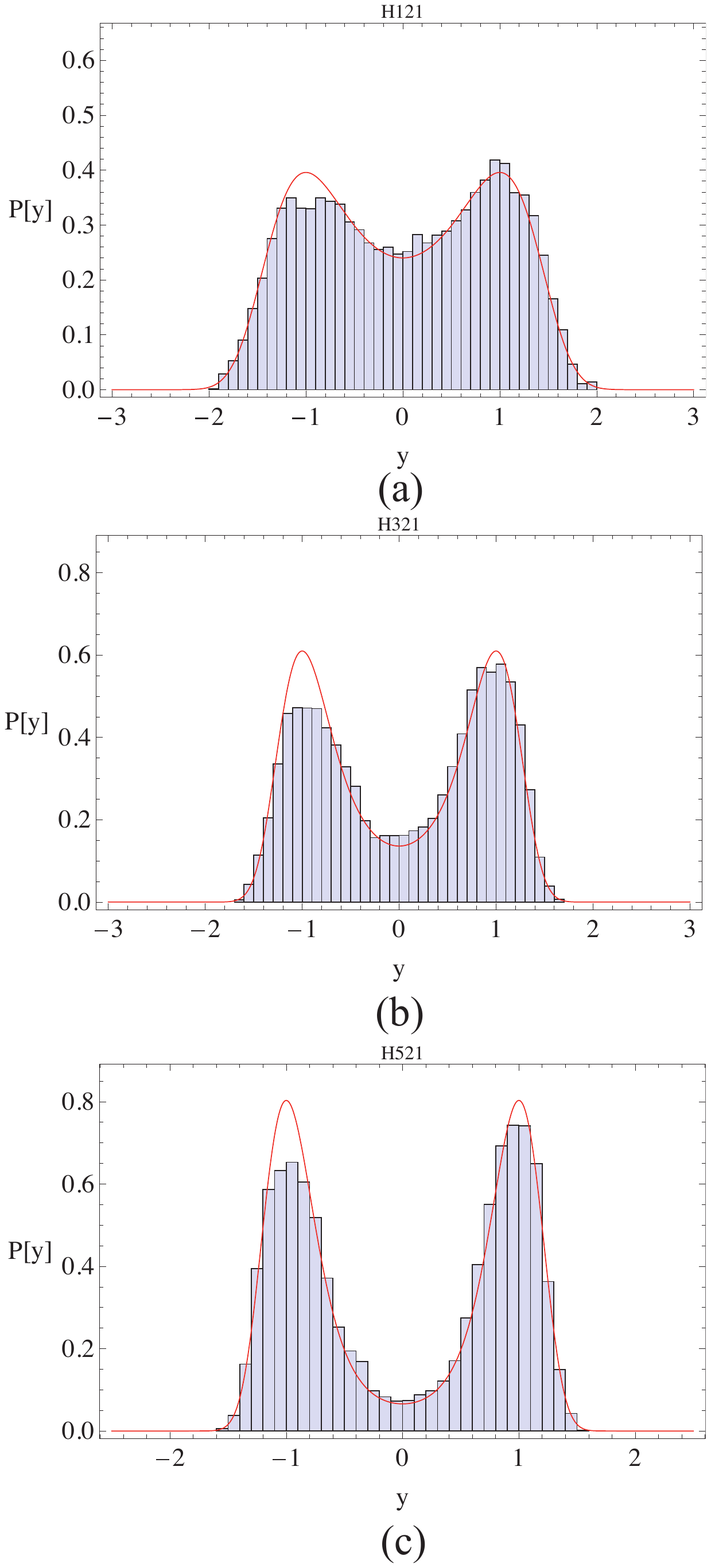}\\
 \end{center}
 \end{figure}

   \vspace*{1.5cm}
   FIGURE 5

\newpage

\begin{figure}[htb!]
 \begin{center}
 \includegraphics[width=3.5in]{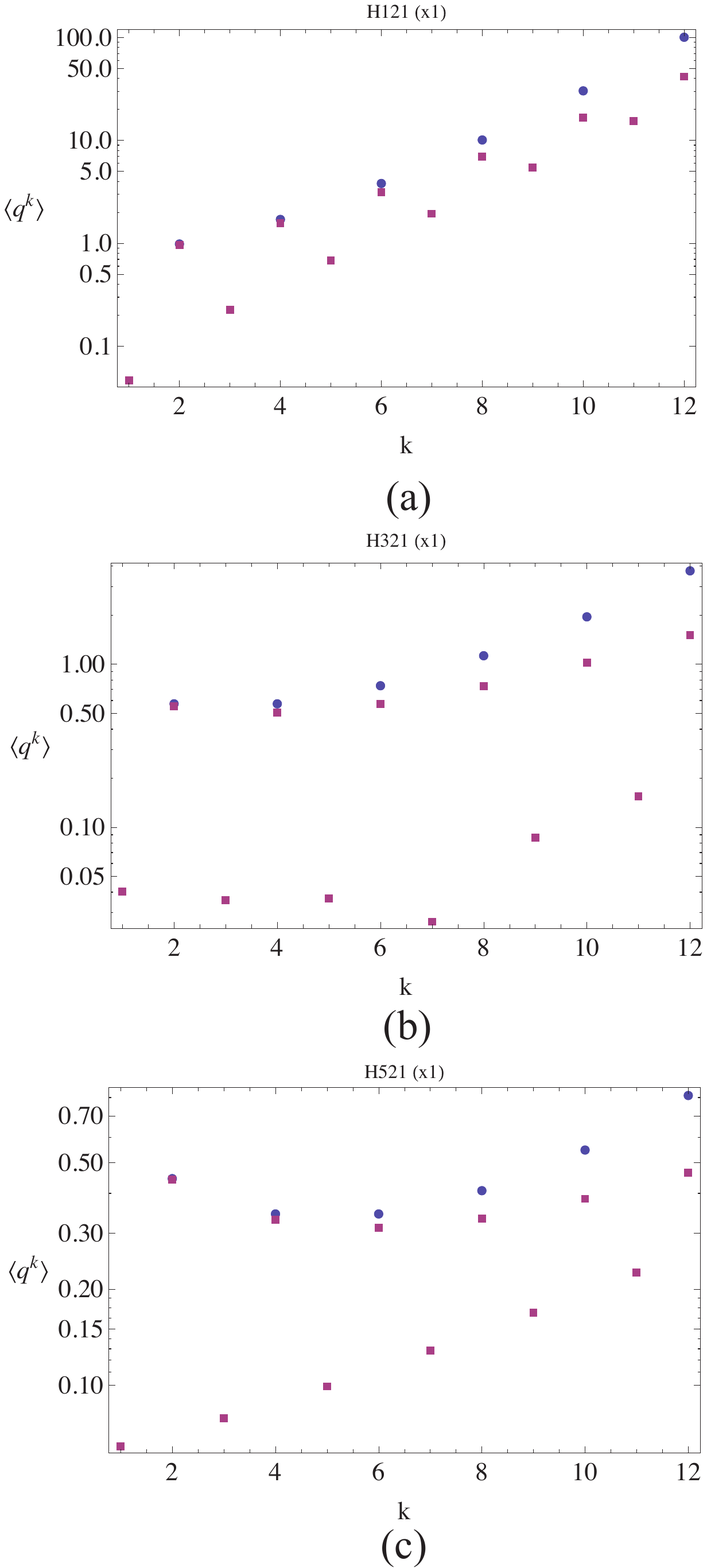}
 \end{center}
\end{figure}

   \vspace*{1.5cm}
   FIGURE 6

\newpage

\begin{figure}[htb!]
 \begin{center}
 \includegraphics[width=3.5in]{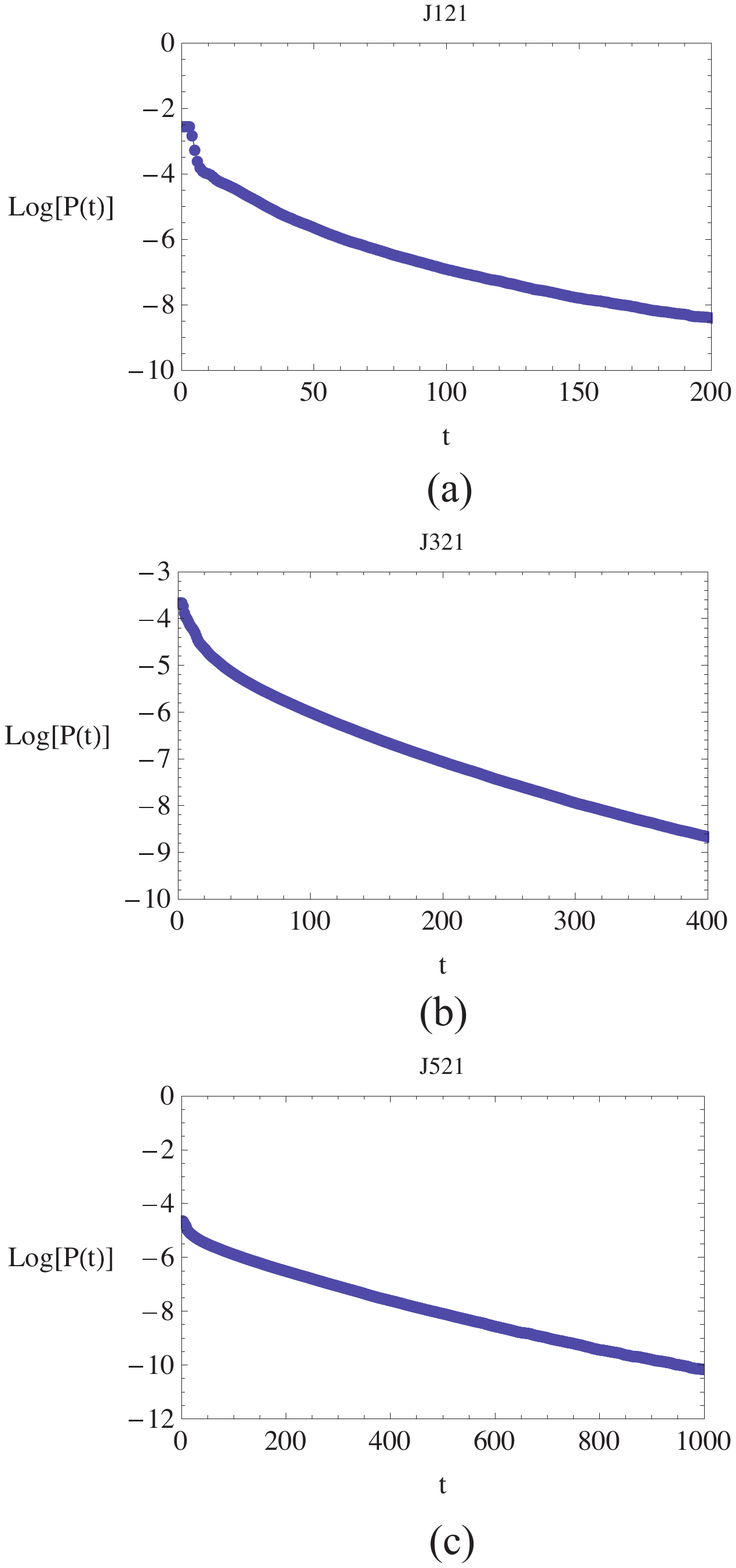}
 \end{center}
 \end{figure}

   \vspace*{1.5cm}
   FIGURE 7

\newpage

\begin{figure}[htb!]
 \begin{center}
 \includegraphics[width=3.5in]{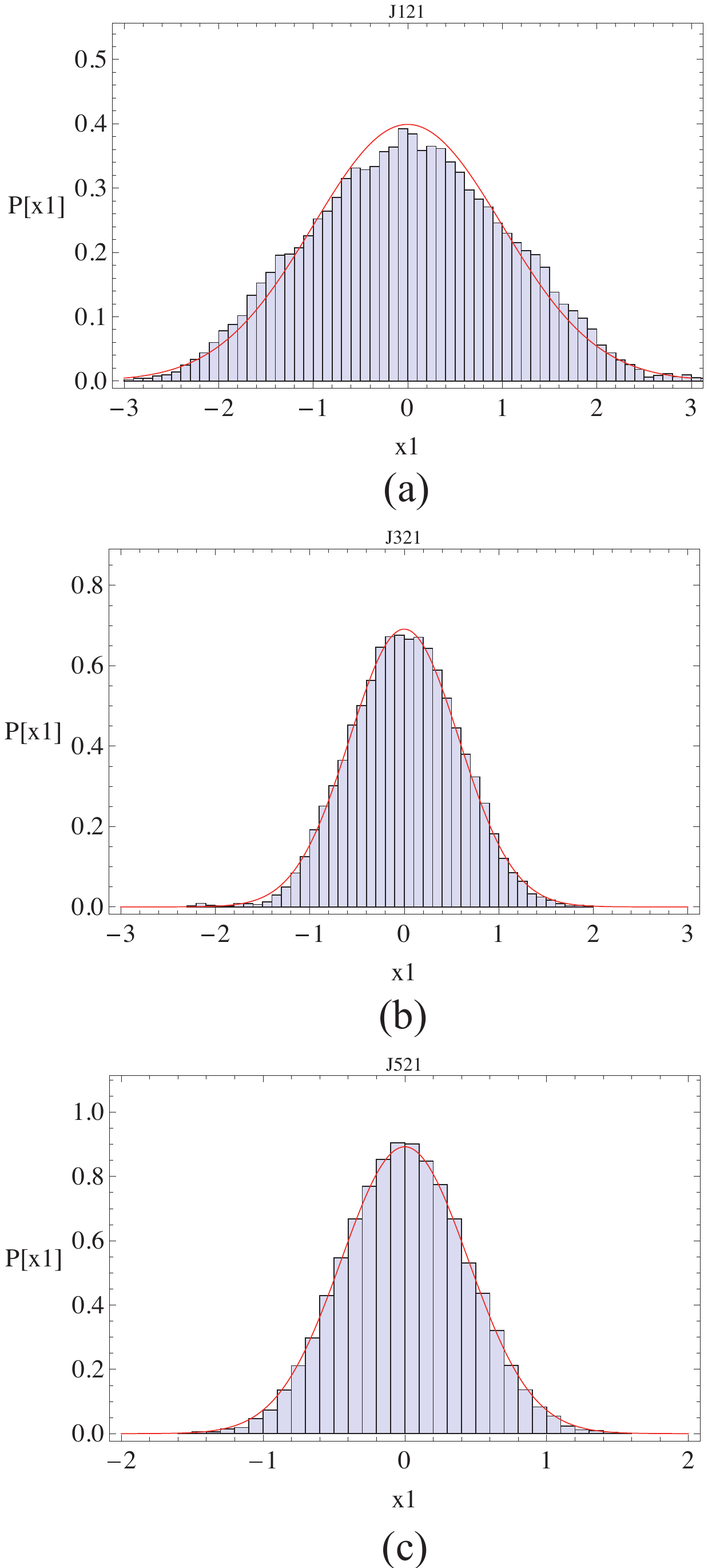}
 \end{center}
\end{figure}

   \vspace*{1.5cm}
   FIGURE 8

\newpage

\begin{figure}[htb!]
 \begin{center}
 \includegraphics[width=3.5in]{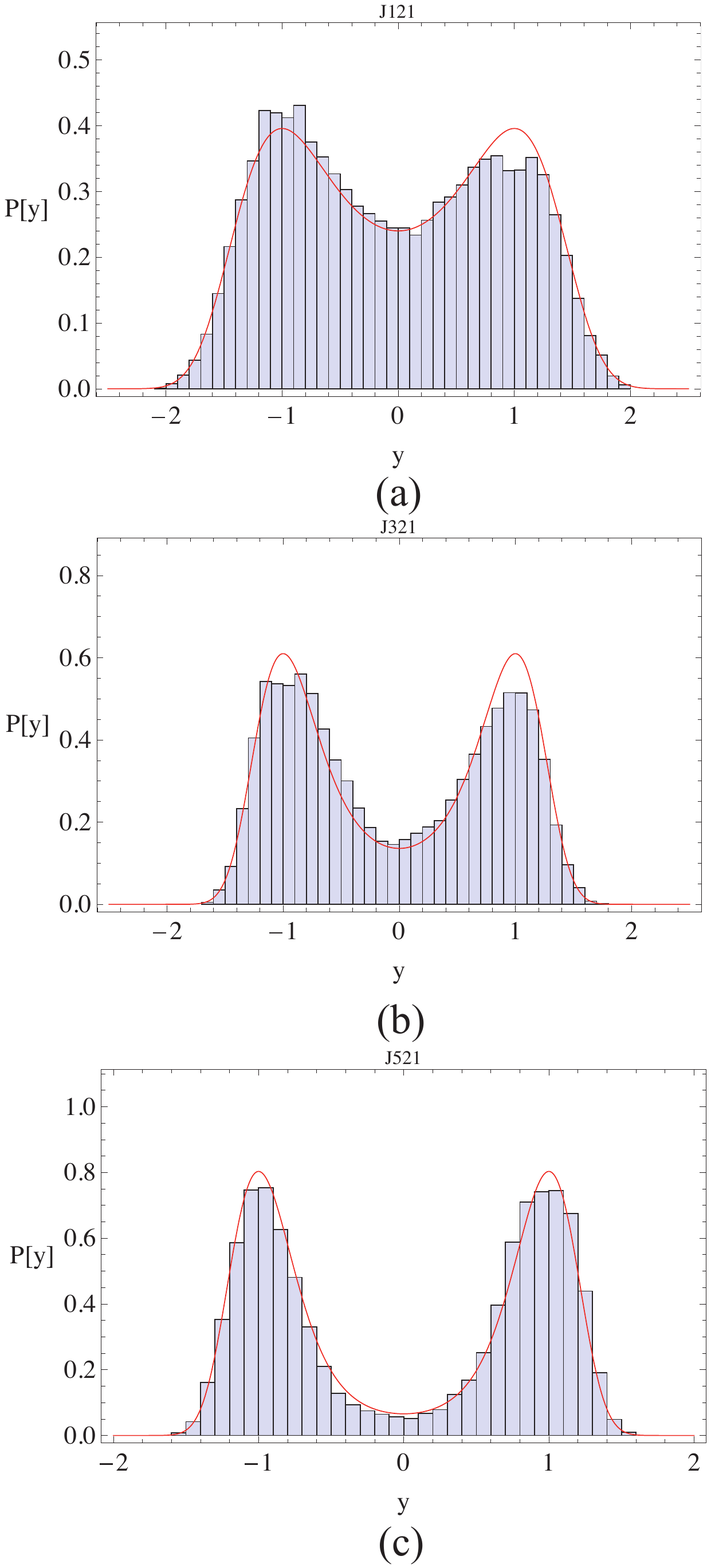}
 \end{center}
 \end{figure}

   \vspace*{1.5cm}
   FIGURE 9

\newpage

\begin{figure}[htb!]
 \begin{center}
 \includegraphics[width=3.5in]{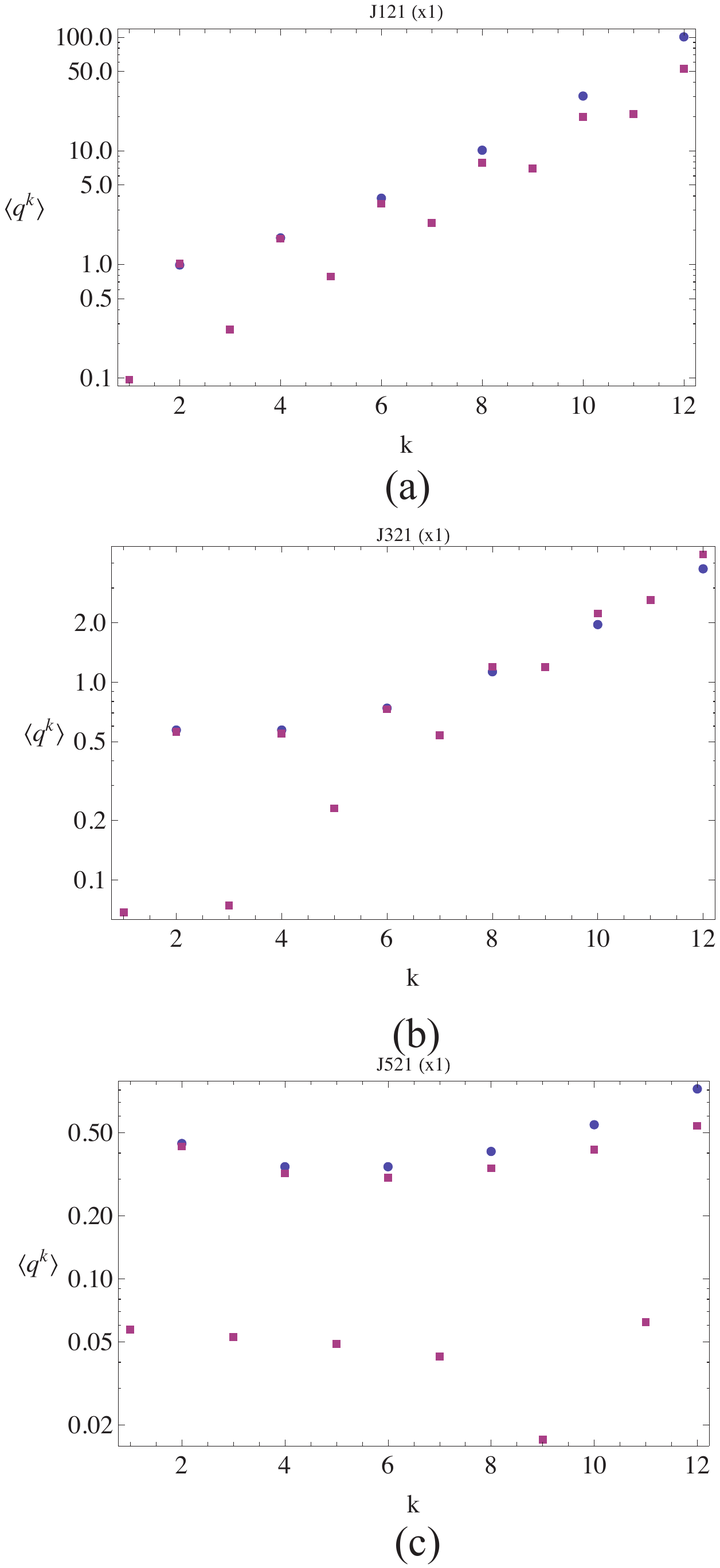}
 \end{center}
\end{figure}

   \vspace*{1.5cm}
   FIGURE 10

\newpage

\begin{figure}[htb!]
 \begin{center}
 \includegraphics[width=4.in]{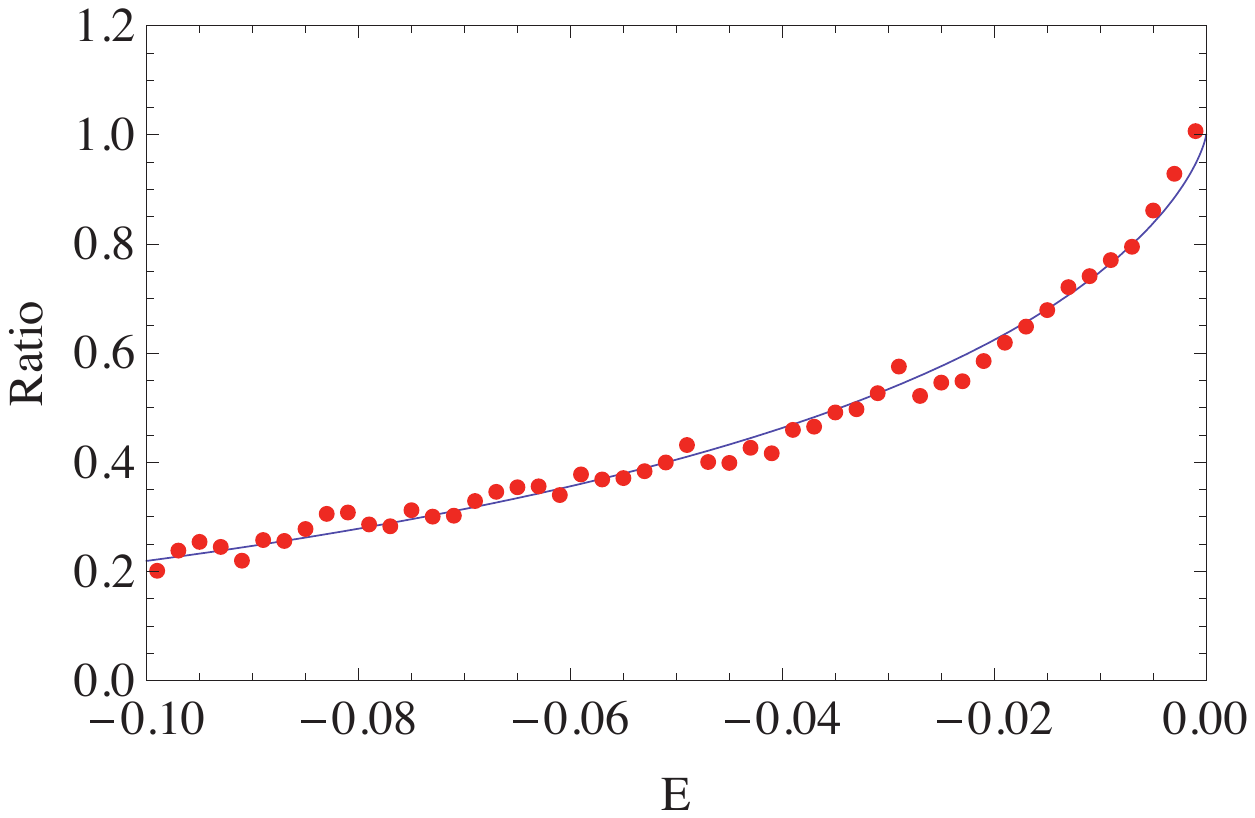}\\
 \end{center}
 \end{figure}

   \vspace*{1.5cm}
   FIGURE 11

\newpage

\begin{figure}[htb!]
 \begin{center}
 \includegraphics[width=4.in]{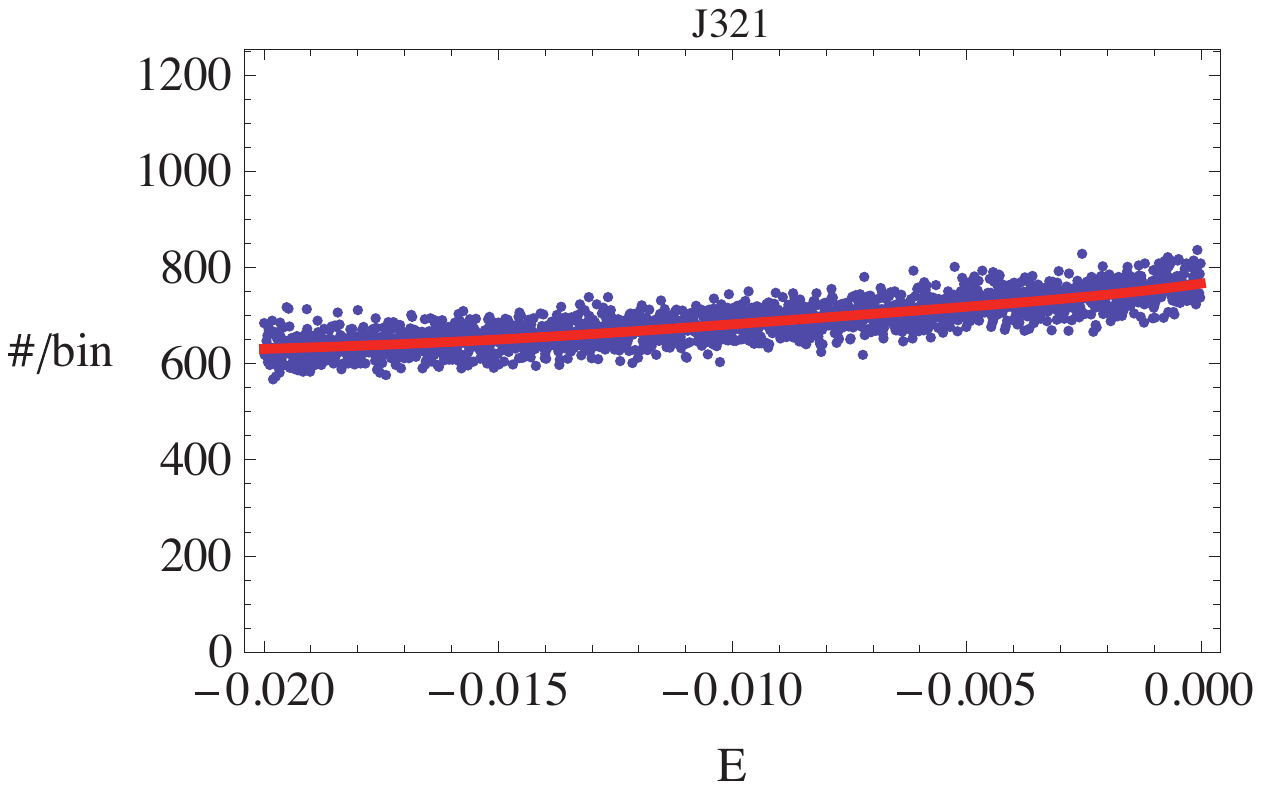}
 \end{center}
\end{figure}

   \vspace*{1.5cm}
   FIGURE 12

\end{document}